\documentclass[11pt,a4paper]{article}
\usepackage{jheppub}
\usepackage{epsf}
\usepackage{amsmath,cases,amssymb,amsfonts,bbold,relsize,exscale,scalefnt,anyfontsize}
\usepackage{graphicx,graphics,epstopdf}
\usepackage{url}
\usepackage{color}
\usepackage{subfig}

\newcommand{\nn}{\nonumber} 
\newcommand{\beq}{\begin{equation}}
\newcommand{\eeq}{\end{equation}} 
\newcommand{\beqa}{\begin{eqnarray}} 
\newcommand{\eeqa}{\end{eqnarray}}

\newcommand{\unit}{1\!\!1}

\definecolor{red}{rgb}{1,0,0}

\def\bc{\begin{center}}
\def\ec{\end{center}}
\def\bi{\begin{itemize}}
\def\ei{\end{itemize}}
\def\be{\begin{equation}}
 \def\ee{\end{equation}}
\def\ben{\begin{equation*}}
 \def\een{\end{equation*}}
 \def\bea{\begin{eqnarray}}
 \def\eea{\end{eqnarray}}
 \def\bean{\begin{eqnarray*}}
 \def\eean{\end{eqnarray*}}
\newcommand{\ie}{{\em i.e.}}  \newcommand{\eg}{{\em e.g.}}

\def\xtwo{x_{_2}}

\newcommand{\lqcd}{\Lambda_{_{\rm QCD}}} 
  
\newcommand{\morder}[1]{{\cal O}\left(#1 \right)}
\newcommand{\eq}[1]{(\ref{#1})}
\newcommand{\ed}{\end{document}}

\newcommand{\ave}[1]{\langle{#1}\rangle}

\newcommand{\tr}{\mathrm{Tr}}

\newcommand{\zerovec}{{\boldsymbol 0}}

\newcommand{\bvec}{{\boldsymbol b}}
\newcommand{\rvec}{{\boldsymbol r}}

\newcommand{\pvec}{{\boldsymbol p}}
\newcommand{\ellvec}{{\boldsymbol \ell}}
\newcommand{\kvec}{{\boldsymbol k}}

\newcommand{\xvec}{{\boldsymbol x}}

\newcommand{\vvec}{{\boldsymbol v}}

\newcommand{\sqrts}{\sqrt{s}}

\newcommand{\jpsi}{{\mathrm J}/\psi}
\newcommand{\xf}{x_{\mathrm{F}}}

\newcommand{\pp}{p--p}
\newcommand{\pA}{p--A}
\newcommand{\pPb}{p--Pb}

\newcommand{\hi}{A--A}
\newcommand{\dd}{{\rm d}}
  
\newcommand{\lsim}{\lesssim} \newcommand{\gsim}{\gtrsim}

\newcommand{\re}{{\,\rm Re \,}}

 \def\esim{\,\mathrel{\rlap{\lower0.2em\hbox{$-$}}\raise0.15em\hbox{\footnotesize $\hskip0.04em\sim$}}\,}
 \def\gsim{\mathrel{\rlap{\lower0.2em\hbox{$\sim$}}\raise0.2em\hbox{$>$}}}
 \def\ksim{\mathrel{\rlap{\lower0.2em\hbox{$\sim$}}\raise0.2em\hbox{$<$}}}

\title{Medium-induced gluon radiation in hard forward parton scattering in the saturation formalism}

\author[a]{St\'ephane Munier,}
\author[b]{St\'ephane Peign\'e,}
\author[a,c]{Elena Petreska}

\affiliation[a]{Centre de Physique Th\'eorique, Ecole Polytechnique, CNRS, Universit\'e Paris-Saclay, \\ 91128 Palaiseau, France}
\affiliation[b]{SUBATECH, UMR 6457, Universit\'e de Nantes, Ecole des Mines de Nantes, CNRS/IN2P3, \\ 4 rue Alfred Kastler, 44307 Nantes cedex 3, France}
\affiliation[c]{Departamento de F\'isica de Part\'iculas and IGFAE, Universidade de Santiago de Compostela, 15782 Santiago de Compostela, Spain}

\emailAdd{stephane.munier@polytechnique.edu} \emailAdd{peigne@subatech.in2p3.fr} \emailAdd{elena.petreska@usc.es}

\abstract{We derive the medium-induced, coherent gluon radiation spectrum associated with the hard forward scattering of an energetic parton off a nucleus, in the saturation formalism and within the Gaussian approximation for the relevant correlators of Wilson lines. The calculation reproduces the simple expression for the spectrum previously obtained in the opacity expansion formalism, and rigorously specifies its validity range. The connection between the calculations in the opacity expansion and saturation formalisms is made apparent. This study may serve as a first step in order to implement consistently induced coherent energy loss and gluon shadowing in `saturation-based models' of hadron nuclear suppression in proton-nucleus collisions.}


\begin{document}

\maketitle
\setcounter{footnote}{0}
\renewcommand{\thefootnote}{\arabic{footnote}}


\section{Introduction}

At collider (RHIC, LHC) energies and at large enough rapidity, the observed suppression of hadron production rates in proton-nucleus (\pA) with respect to proton-proton (\pp) collisions, is often attributed to gluon `shadowing', \ie\ to the depletion of the gluon density in the nucleus (with respect to a proton) expected at small $\xtwo \lesssim 10^{-2}$. Shadowing is currently either incorporated in nuclear parton distribution functions (nPDFs) obtained from fits based on DGLAP evolution within collinear factorization, or determined from non-linear QCD evolution within the saturation formalism (see~\cite{Armesto:2006ph,Gelis:2010nm} for topical reviews).

In addition to shadowing, another effect can also suppress hadron production rates, namely medium-induced coherent radiation in cold nuclear matter~\cite{Arleo:2010rb,Arleo:2012rs,Armesto:2012qa,Arleo:2013zua,Armesto:2013fca,Peigne:2014uha,Liou:2014rha,Peigne:2014rka}. As a matter of fact, in quarkonium production the effect of coherent energy loss was shown~\cite{Arleo:2012rs,Arleo:2013zua} to provide a good description of fixed-target~\cite{Leitch:1999ea} and RHIC~\cite{Adare:2010fn,Adare:2012qf} $\jpsi$ nuclear suppression data, as well as successful predictions for the $\jpsi$ suppression later observed in \pPb\ collisions at the LHC~\cite{Abelev:2013yxa,Aaij:2013zxa}. Moreover, when extrapolated to nucleus-nucleus (\hi) collisions, coherent energy loss in cold nuclear matter proves to be a sizable effect~\cite{Arleo:2014oha}, which is therefore crucial to study in more detail before disentangling the additional `hot' effects associated with the production of a quark-gluon plasma in \hi\ collisions. 

Shadowing and coherent energy loss are two different effects, and should in principle be both taken into account in nuclear suppression models. However, recent models for quarkonium nuclear suppression either focus on shadowing effects (in the collinear factorization approach~\cite{Albacete:2013ei,Vogt:2015uba,Ferreiro:2013pua} or in the saturation formalism~\cite{Fujii:2013gxa,Ducloue:2015gfa,Ma:2015sia}) and neglect energy loss, or consider coherent energy loss as the leading effect to which shadowing is added by assuming that the two effects factorize~\cite{Arleo:2012rs,Arleo:2013zua}. In order to better quantify the respective roles of shadowing and energy loss in quarkonium (and more generally hadron) nuclear suppression in \pA \ collisions at collider energies,\footnote{At fixed-target energies ($\sqrts \lsim 40$~GeV) the situation for $\jpsi$ nuclear suppression seems relatively clear: shadowing is expected to be small at those energies, and coherent energy loss is likely to be the dominant effect, at least at large enough $\xf$~\cite{Arleo:2012rs,Arleo:2013zua}.} a phenomenological approach may consist in looking for specific observables sensitive to energy loss, but having a reduced uncertainty on the precise magnitude of shadowing. For instance, the {\it ratio} of \pA\ nuclear suppression factors in $\jpsi$ w.r.t. Drell-Yan production could be such an observable~\cite{Arleo:2015qiv}. On the theoretical side, a first principle calculation incorporating consistently the physics of both shadowing and coherent energy loss would be desirable but is still missing.

The saturation formalism should be an appropriate framework to achieve this goal. However, to our knowledge only one study addressed the theoretical calculation of induced coherent radiation in this formalism (Ref.~\cite{Liou:2014rha}, see below), and as mentioned above, coherent energy loss has not yet been incorporated in phenomenological studies of hadron nuclear suppression based on saturation. Roughly speaking, current saturation models account for the effects arising directly from multiple soft scatterings in the nucleus, which include small-$x$ shadowing, but also the effect of nuclear transverse momentum broadening (the `Cronin effect'), $\Delta p_\perp^2 \sim Q_s^2$, where $Q_s \equiv Q_g$ is the gluon saturation scale in the nucleus.\footnote{\label{foot:sat-def}Throughout this paper we denote the `saturation scale' of a parton $a=q,g$ in the nucleus and in the proton as $Q_{a}$ and $Q_{a \rm p}$, respectively, and define the generic saturation scale in the nucleus as $Q_s \equiv Q_g$.} Medium-induced coherent energy loss may naively be viewed as `next-to-leading' (since it is formally $\sim \morder{\alpha_s}$ as compared \eg\ to the Cronin effect), but could however be a crucial effect in hadron nuclear suppression (as suggested by $\jpsi$ suppression in \pA\ collisions); its specific parametric dependence indeed leads to an average induced energy loss proportional to the energy of the fast radiating parton, hence its expected relevance in forward processes~\cite{Arleo:2010rb}. 

Medium-induced coherent radiation has previously been studied theoretically by se\-ve\-ral groups, using different formalisms and considering different particular cases. Ref.~\cite{Arleo:2010rb} (see also \cite{Arleo:2012rs}) studied the induced coherent radiation associated with the hard forward $g \to Q\bar{Q}$ process (mediated by a single hard gluon exchange in the $t$-channel), the final $Q\bar{Q}$ pair being a massive {\it pointlike} color octet, in a Feynman diagram calculation and at first order in the opacity expansion formalism~\cite{Gyulassy:2000er}. Induced coherent radiation was also studied using a semi-classical method in Refs.~\cite{Armesto:2012qa,Armesto:2013fca}, at first order~\cite{Armesto:2012qa} and all orders~\cite{Armesto:2013fca} in opacity and in a similar kinematical setup as that of Ref.~\cite{Arleo:2010rb}, however for the $q \to q$ case of a massless quark experiencing a hard scattering mediated by a {\it color singlet} exchange in the $t$-channel. In Ref.~\cite{Peigne:2014uha}, induced coherent radiation was revisited and derived to all orders in the opacity expansion, for any $1\to 1$ hard forward process. A general expression for the induced coherent spectrum (encompassing the particular cases studied before) was found, as well as a simple rule for the overall color factor of the induced spectrum. The latter reads $C_R + C_{R'} - C_{t}$, with $C_R$, $C_{R'}$ and $C_t$ the incoming, outgoing and $t$-channel color charges of the $1\to 1$ hard partonic process~\cite{Peigne:2014uha}. Induced coherent radiation associated with $1 \to 2$ forward processes was also addressed (in the leading-logarithm and large-$N_c$ limits), for $g \to q \bar q$ and $q \to q g$ in the saturation formalism~\cite{Liou:2014rha} and for $q \to q g$ and $g \to gg$ using the opacity expansion~\cite{Peigne:2014rka}. In the leading-logarithm approximation, the soft induced coherent radiation does not probe the size of the final two-parton system, and thus only depends on its {\it total} color charge. Hence the conjecture, proposed and explicitly checked for $q \to q g$ and $g \to gg$ in Ref.~\cite{Peigne:2014rka}, that the spectrum associated with $1 \to n$ hard forward processes is given by an incoherent sum of spectra associated with $1 \to 1$ processes, weighted by the probabilities $P_{R'}$ for the $n$-parton state to be produced in color representation $R'$ in the hard process. This conjecture is expected to hold for any finite $N_c$ (but only in the leading-logarithm approximation)~\cite{Peigne:2014rka}. 

In view of implementing coherent energy loss in saturation-based phenomenological models, it is helpful to first show how the induced coherent spectrum associated with $1 \to 1$ hard forward processes, previously derived in Ref.~\cite{Peigne:2014uha} using the opacity expansion, arises in the saturation formalism. This is the goal of the present study, which may also serve as a first step in order to explore, in the saturation formalism, the above-mentioned conjecture. We restrict ourselves to $1 \to 1$ processes where the type $a$ of the energetic parton is conserved, namely $a \to a$, with $a=q$ (massless quark) or $a=g$ (gluon).\footnote{From this point of view, the present calculation is less general than in Ref.~\cite{Peigne:2014uha}, where hard processes with a change of the energetic parton type, $q \to g$ and $g \to q$, were also considered. Such processes could certainly be studied within the saturation formalism, but this would require implementing in this formalism hard scatterings with {\it color triplet} exchange in the $t$-channel.} As in Ref.~\cite{Peigne:2014uha} we work at finite $N_c$ and at `leading order' in $\alpha_s$, \ie\ we consider the radiation of a {\it single} gluon. Our final result \eq{spec-master-pointlike} for the induced coherent spectrum associated with $a \to a$ coincides with the result of Ref.~\cite{Peigne:2014uha}, in particular the color factor $2 C_R -C_A$ obeys the general rule $C_R + C_{R'} - C_{t}$ ($a \to a$ being mediated by a $t$-channel {\it color octet} hard exchange, $C_t = C_A$). In the quark case we have $2 C_F -C_A = -1/N_c$, and the ({\it induced}) radiation spectrum associated with $q \to q$ is thus negative. This somewhat surprising result (which however has a simple interpretation~\cite{Peigne:2014uha}) is confirmed here within the saturation formalism. 

Finally, in the present study we also evaluate the magnitude of expected corrections to the spectrum \eq{spec-master-pointlike}, allowing one to specify its validity range, which question was not addressed in Ref.~\cite{Peigne:2014uha}. We find that \eq{spec-master-pointlike} holds up to values of $z$ (the energy fraction carried by the radiated gluon w.r.t. to the parton $a$) where \eq{spec-master-pointlike} is much smaller than its leading logarithmic behavior $\sim \log{(1/z)}$ valid at small $z$, but still larger in order of magnitude than the expected corrections. Thus, \eq{spec-master-pointlike} predicts the `large'-$z$ tail of the spectrum, which is important for phenomenology. 

Our paper is organized as follows. 

In section~\ref{sec:induced}, we formulate our observable in coordinate space and arrive at the general expression of the induced coherent radiation spectrum, see Eqs.~\eq{spec-ind-1}-\eq{spec-ind-INT}. Within the Gaussian approximation for the field of the nucleus, we extract in section~\ref{sec:eval} the leading asymptotics of the coherent spectrum, given by \eq{spec-master-pointlike}. For simpli\-ci\-ty, the calculation of section~\ref{sec:eval} is somewhat formal, but all details of the fully rigorous calculation are given in appendix~\ref{app:detailed}. 
In section~\ref{sec:comp-form} we repeat the derivation of \eq{spec-master-pointlike} in a simplified and transparent way, allowing one to connect our calculation performed in the saturation formalism to the opacity expansion of Ref.~\cite{Peigne:2014uha}. Finally, in section~\ref{sec:smallx} we discuss heuristically how small-$x$ evolution could be implemented in the calculation of induced coherent energy loss.  

\section{Medium-induced soft radiation in hard forward parton scattering}
\label{sec:induced} 

\subsection{Definition of observable}

We consider an energetic massless parton $a=q, g$ (massless quark or gluon) from a projectile proton that scatters off a target nucleus A, and later hadronizes in the forward region of the proton-nucleus (\pA) collision, $a(P_0) +  {\rm A} \to a({P}) + {\rm X}$.\footnote{We will use light-cone coordinates, $P=(p^+,p^-, \pvec)$ with $p^{\pm} = (p^0 \pm p^z)/\sqrt{2}$, and choose the proton and nucleus to be moving in the $+$ and $-$ directions, respectively. The forward region is defined as the proton fragmentation region, \ie\ the region of positive rapidities in the center-of-mass frame of an elementary proton-nucleon collision, corresponding to large rapidities in the nucleus rest frame. We will set $\pvec_{0} = \zerovec$ and denote $p \equiv |\pvec|$.} We consider the limit of {\it forward} scattering, where $p^+$ is arbitrarily large but transverse momenta are limited, and focus on  {\it hard} scattering, where the final parton transverse momentum $p \equiv |\pvec|$ is much larger than the gluon saturation scale $Q_s$ in the nucleus,
\be
\label{hard-limit}
p \gg Q_s \, .
\ee

We aim to derive the {\it medium-induced}, {\it soft} gluon radiation spectrum associated with such a scattering of parton $a$. This amounts to producing an additional gluon in the final state, $a(P_0) +  {\rm A} \to a({P)} + g({P'}) + {\rm X}$, and considering the limit 
\be
\label{soft-limit}
z \equiv \frac{p'^+}{p_0^+} \ll 1  \, .
\ee
The soft radiation energy spectrum is then defined as
\be 
\label{spec-pA}
z \frac{dI_{\rm A}}{dz} =  p'^+ \frac{dI_{\rm A}}{dp'^+} = \mbox{\fontsize{18}{2} \selectfont {$\frac{\ \ \frac{d\sigma(a+{\rm A} \to a+ g + {\rm X})}{dy\, dy' d^2 \pvec} \ \ }{\ \ \frac{d\sigma(a +{\rm A} \to a + {\rm X})}{dy \, d^2 \pvec}\ \ } $}}=  \mbox{\fontsize{18}{2} \selectfont {$\frac{\ \ \frac{d\sigma({\rm p} +{\rm A} \to a + g + {\rm X})}{dy\, dy' d^2 \pvec}\ \ }{\ \ \frac{d\sigma({\rm p} +{\rm A} \to a + {\rm X})}{dy \, d^2 \pvec}\ \ }$}} \, ,
\ee
where $y = \ln{({p^+}/{p})}$ is the parton rapidity, and $d\sigma({\rm p} +{\rm A} \to \ldots)$ is obtained by convoluting $d\sigma(a + {\rm A} \to \ldots)$ with the distribution of parton $a$ in the proton $f_{a/{\rm p}}(x_{\rm p}, \mu_{_{\rm F}}^2)$, with $x_{\rm p} = p_0^+/p_{\rm p}^+$ ($p_{\rm p}^+$ is the longitudinal momentum of the proton) and $\mu_{_{\rm F}}$ the factorization scale. (For our purpose we do not need to implement fragmentation functions of the final partons into hadrons.)

The {\it medium-induced} spectrum is given by the difference of \eq{spec-pA} between a nucleus and a proton target,
\be
\label{ind-prescription}
\left. z \frac{dI}{dz} \right|_{\rm ind}  \equiv z \frac{dI_{\rm A}}{dz} - z \frac{dI_{\rm p}}{dz} \, .
\ee
The difference arises only from the different proton and nucleus saturation scales.

\subsection{Unified formulae for $a \to a + g$ forward production}
\label{sec:prelim}

The forward production cross section $d\sigma({\rm p}{\rm A} \to a g + {\rm X})$ appearing in the numerator of \eq{spec-pA} has been considered before by various authors. We start by quoting the double differential cross section derived in the case $a=q$~\cite{Marquet:2007vb} and $a=g$~\cite{Iancu:2013dta} for an arbitrary fraction  $z$ of the final gluon energy.\footnote{We assume $z \sim \morder{1}$ in the present section and will take the soft limit $z \ll 1$ later in section~\ref{sec:master}.} The cross section is given by the same formal expression in the two cases and reads 
\bea
&& \hspace{-20pt}\frac{d\sigma({\rm pA}\to ag + {\rm X})}{dy \, dy' d^2 \pvec \,d^2 \pvec'} = \frac{\alpha_s}{(2\pi)^6} \, x_{\rm p} f_{a/{\rm p}}(x_{\rm p},\mu_{_{\rm F}}^2) \, z \, (1-z) \,  \Phi_a^g(z) \nn \\ 
&& \times \int_{\xvec} \, \int_{\xvec'} \, \int_{\bvec} \, \int_{\bvec'}  \, e^{-i \pvec \cdot(\bvec-\bvec')-i \pvec' \cdot(\xvec-\xvec')} \, \frac{({\xvec} - {\bvec}) \cdot ({\xvec'} - {\bvec'})}{({\xvec} - {\bvec})^2 \, ({\xvec'} - {\bvec'})^2}  \,  \nonumber \\
&& \times \left\{S^{(4)}_{a}[{\bvec},{\xvec},{\bvec'},{\xvec'}] + S^{(2)}_{a}[\vvec,\vvec'] - S^{(3)}_{a}[{\bvec},{\xvec},\vvec'] - S^{(3)}_{a}[\bvec',{\xvec'},{\vvec}]^{*} \right\}\, ,
\label{a2ag-cross}
\eea
where we use the notation $\int_{\xvec} \equiv \int d^2\xvec$, $S^{(i)}_a$ are correlators of Wilson lines (see below), and $\Phi_a^g(z)$ is the $q\to g$ ($a=q$) or $g\to g$ ($a=g$) splitting function given by~\cite{Dokshitzer:1991wu}
\be
\label{splittings}
\Phi_q^g(z) = C_F \times  2 \frac{1 + (1-z)^2}{z} \ ; \ \ \Phi_g^g(z) = C_A \times 4 \left[\frac{z}{1-z}+\frac{1-z}{z}+z(1-z)\right] \, .
\ee

\begin{figure}
\centering
\subfloat[]{\includegraphics[width=0.4\textwidth]{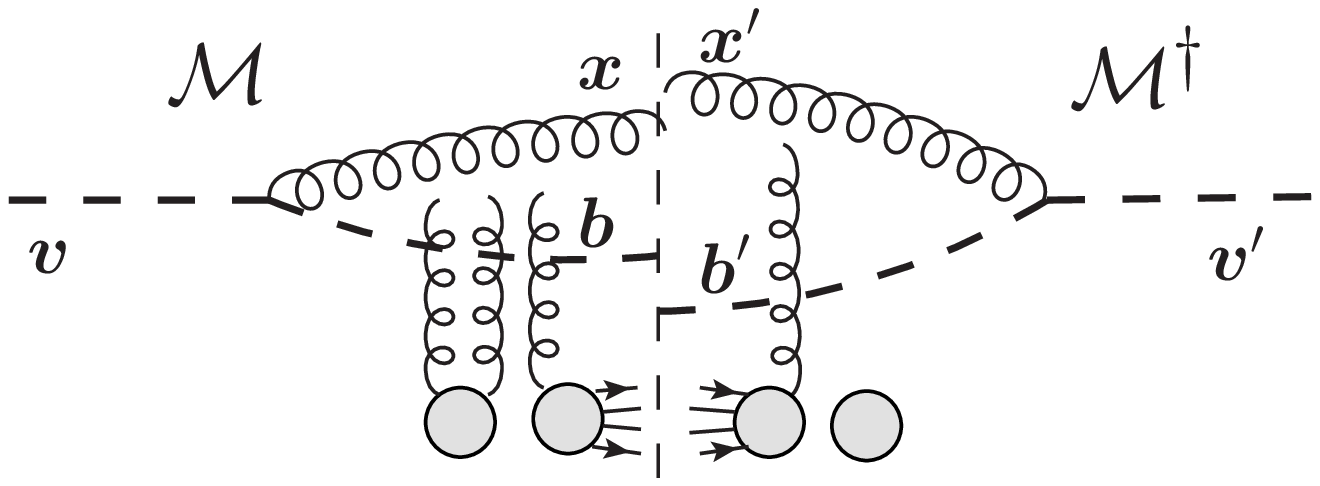}\label{fig:initial}}
\hskip 15mm 
\subfloat[]{\includegraphics[width=0.4\textwidth]{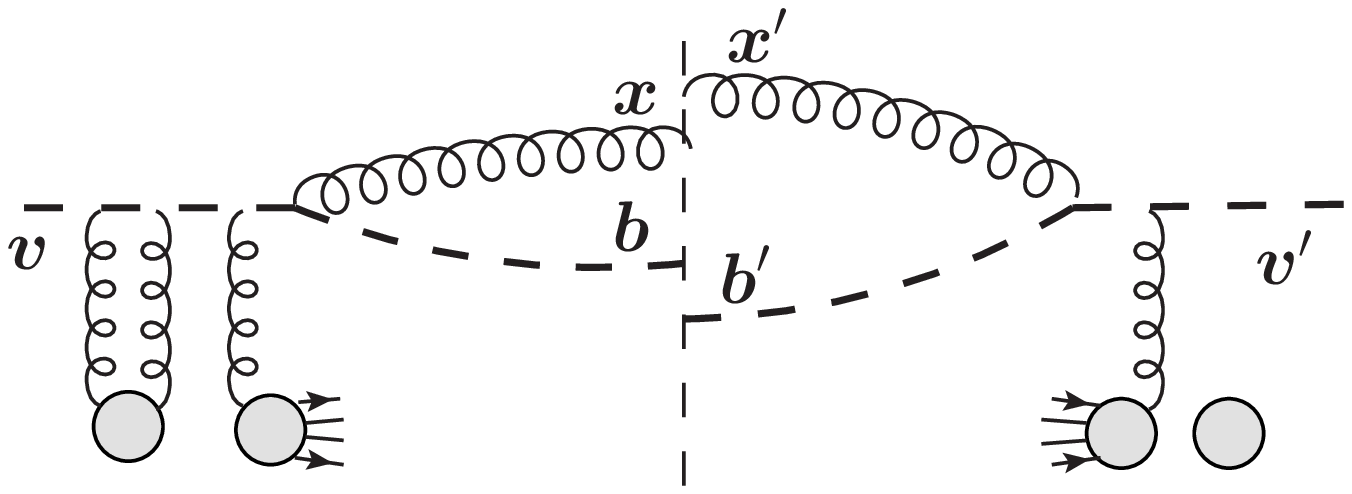}\label{fig:final}} \\ 
\subfloat[]{\includegraphics[width=0.4\textwidth]{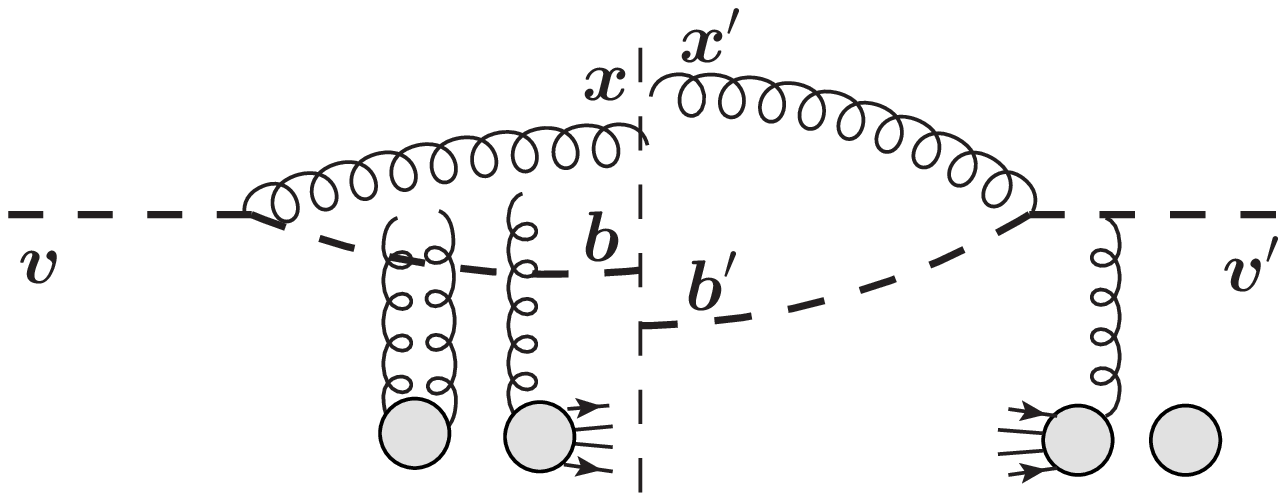}\label{fig:interferenceIF}} 
\hskip 15mm 
\subfloat[]{\includegraphics[width=0.4\textwidth]{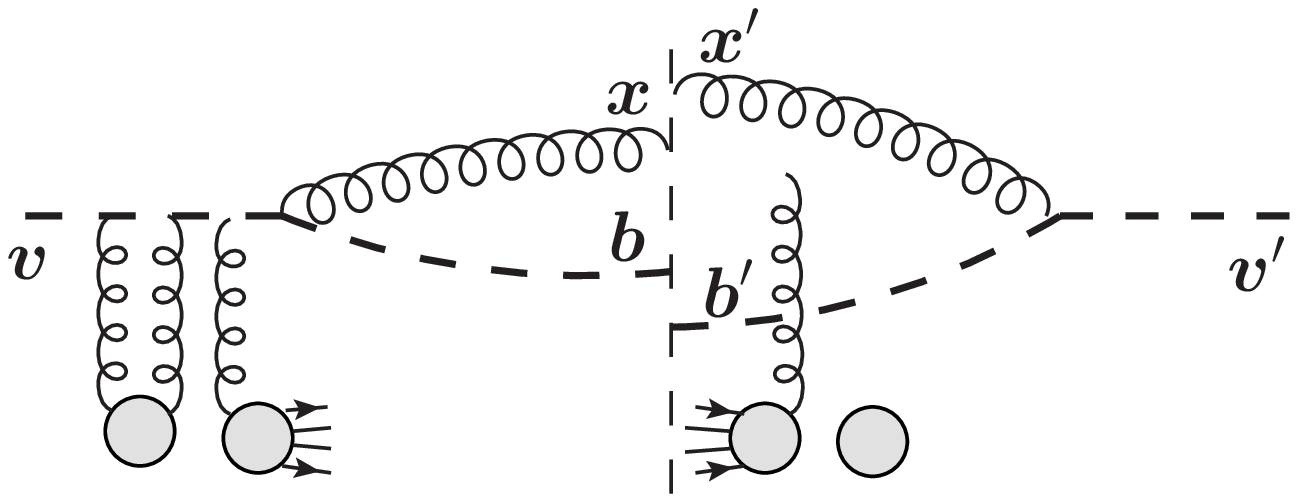}\label{fig:interferenceFI}}
\caption{Diagrams contributing to the cross section for $a+ {\rm A} \to {a} + g + {\rm X}$, where the parton $a=q, g$ is represented by the dashed line. The scattering off the nucleus consists in a number of two-gluon exchanges between the $(a,g)$ system and a subset of nucleons in the nucleus. We represented a particular event where a first nucleon interacts elastically in the amplitude and does not interact in the conjugate amplitude, and a second nucleon that interacts inelastically. The parton $a$ can radiate the gluon from the initial state (diagram (a)), the final state (diagram (b)), or coherently from the interference between initial-state and final-state emissions (diagrams ({c}) and (d)).} 
\label{fig:FIG1}
\end{figure}

The four terms in the bracket of \eq{a2ag-cross} arise from the four diagrams contributing to the squared amplitude of the partonic process $a(P_0)+ {\rm A} \to {a}({P}) + g(P') + {\rm X}$ and are represented in Fig.~\ref{fig:FIG1}: diagram~\ref{fig:initial} where the gluon is radiated before the interaction with the nucleus in both the amplitude and its conjugate (initial state radiation); diagram~\ref{fig:final} where the splitting $a \to a g$ happens after the interaction in both the amplitude and its conjugate (final state radiation); diagrams~\ref{fig:interferenceIF} and~\ref{fig:interferenceFI} corresponding to interference terms. The coordinates of the outgoing parton $a$ and gluon are, respectively, $\bvec$ and ${\xvec}$ in the amplitude, and $\bvec'$ and $\xvec'$ in the conjugate amplitude. The coordinate of the incoming parton $a$ is ${\vvec}=z{\xvec} +(1-z){\bvec}$ in the amplitude, and ${\vvec'}=z{\xvec'} +(1-z){\bvec'}$ in its conjugate.

Finally, $S^{(i)}$ are correlators of Wilson lines which sum up the multiple scattering interactions of the parton and gluon with the target. The correlators relevant to the diagrams of Figs.~\ref{fig:initial}, \ref{fig:final}, \ref{fig:interferenceIF} and \ref{fig:interferenceFI} read, respectively~\cite{Marquet:2007vb,Iancu:2013dta} 
\bea
S^{(4)}_{a}[{\bvec},{\xvec},{\bvec'},{\xvec'}]&=&\frac{1}{d_R  C_R}\left<\tr \left(T_R^c \, {U_R}({\bvec}) \, {U_R}^\dagger({\bvec'}) \, T_R^d \right)[{V}({\xvec}){V}^\dagger({\xvec'})]^{cd}\right> \, , \label{S4-corr} \\
{S}_a^{(2)}[{\vvec},{\vvec'}]&=&\frac{1}{d_R}\left<\tr \left({U_R}({\vvec}){U_R}^\dagger({\vvec'})\right)\right> \, , \label{S2-corr}\\ 
S^{(3)}_{a}[{\bvec},{\xvec},{\vvec'}] &=& \frac{1}{d_R C_R}\left<\tr \left(T_R^c \, {U_R}({\bvec}) \, T_R^d \, {U_R}^\dagger({\vvec'})\right) \left[ {V}({\xvec}) \right]^{cd} \right> \, ,  \label{S3-corr-1} \\
S^{(3)}_{a}[{\bvec'},{\xvec'},{\vvec}]^{*} &=& \frac{1}{d_R C_R}\left<\tr \left({U_R}({\vvec}) \, T_R^c \, {U_R}^\dagger({\bvec'}) \,T_R^d \right) [{V}^\dagger({\xvec'})]^{cd} \right> \label{S3-corr-2} \, ,
\eea
where the average stands for the average over the gluon field configurations in the nucleus, and $R$ denotes the ${\rm SU}(N_c)$ color representation of parton $a$, namely the fundamental $R= F$ ($a=q$) or adjoint $R= A$ ($a=g$) representation, with corresponding color generators $T_R^c = T_F^c \equiv t^c$ and $T_R^c = T_A^c \equiv T^c$. The quantity $U_R$ is the Wilson line of parton $a$ in light-cone gauge $A^+=0$, 
\be 
{U_R}({\xvec}) = \mathcal{P} \exp \left[ ig \int dx^+ A_c^-(x^+, {\xvec}) \, T_R^c \right] \, ,
\ee
and $V({\xvec}) \equiv {U_A}({\xvec})$. Note that the correlators $S^{(i)}$ are defined so that $S^{(i)} \to 1$ for a zero background field.\footnote{To check this, use $T_R^c T_R^c = C_R \, \unit_R$ and $\tr \, \unit_R = d_R$ in Eqs.~\eq{S4-corr} to \eq{S3-corr-2}, where $C_F = (N_c^2-1)/(2N_c)$, $C_A=N_c$, $d_F = N_c$, $d_A = N_c^2-1$.} It is  worth recalling the identity
\be
\label{adjoint-identity}
[{V}({\xvec})]^{cd} = 2 \,\tr \left( {U_F}^\dagger({\xvec}) t^c {U_F}({\xvec}) t^d \right) \, ,
\ee
implying that $[{V}({\xvec})]^{cd}$ is real, $[{V}^{*}({\xvec})]^{cd} = [{V}({\xvec})]^{cd}$. From this one verifies that \eq{S3-corr-1} and \eq{S3-corr-2} are indeed related by complex conjugation.\footnote{In fact, $S^{(3)}_{a}[{\bvec},{\xvec},{\vvec'}]$ is always real in the particular case $a = g$, as can be easily checked using $(T^a)_{bc} = -if_{abc}$ and evaluating the trace in \eq{S3-corr-1}.} 

We stress that in general, the correlators \eq{S4-corr}-\eq{S3-corr-2} depend on the collision energy of the parton-nucleus scattering, through the average over the nucleus field configurations. However, within the Gaussian approximation for the nucleus field which we use in the present study, this energy dependence is absent: all correlators of Wilson lines can be expressed in terms of (energy-independent) {\it two-point} correlators of the form given by the McLerran-Venugopalan (MV) model~\cite{McLerran:1993ni}, see appendix~\ref{app:Sformulae}. How to implement the energy dependence (or small-$x$ evolution) will be briefly discussed in section~\ref{sec:smallx}.

Since the observable under study defined in \eq{spec-pA} and \eq{ind-prescription} is inclusive in the gluon transverse momentum $\pvec'$, we readily integrate \eq{a2ag-cross} over $\pvec'$, which sets $\xvec' = \xvec$ and gives 
\bea
&& \hspace{40pt}\frac{d\sigma({\rm pA}\to ag + {\rm X})}{dy \, dy' d^2 \pvec} = \frac{\alpha_s}{(2\pi)^4} \, x_{\rm p} f_{a/{\rm p}}(x_{\rm p},\mu_{_{\rm F}}^2) \, z \, (1-z) \,  \Phi_a^g(z) \nn \\ 
&& \times \int_{\xvec} \, \int_{\bvec} \, \int_{\bvec'}  \, e^{-i \pvec \cdot(\bvec-\bvec')} \, \frac{({\xvec} - {\bvec}) \cdot ({\xvec} - {\bvec'})}{({\xvec} - {\bvec})^2 \, ({\xvec} - {\bvec'})^2} \left\{S^{(2)}_{a}[{\bvec},{\bvec'}] + S^{(2)}_{a}[{\vvec},{\vvec'}] - 2 \re S^{(3)}_{a}[{\bvec},{\xvec},\vvec'] \right\}\, , \nn \\
\label{numerator-a}
\eea
where now ${\vvec}=z{\xvec} +(1-z){\bvec}$ and ${\vvec'}=z{\xvec} +(1-z){\bvec'}$. We also used $S^{(4)}_{a}[{\bvec},{\xvec},{\bvec'},{\xvec}] = S^{(2)}_{a}[{\bvec},{\bvec'}]$ (which directly follows from \eq{S4-corr}), 
and the fact that the factor in front of the bracket in the integrand of \eq{numerator-a} is symmetric under $\bvec \leftrightarrow \bvec'$ (given that the cross section is obviously an even function of the vector $\pvec$). 

Finally, let us emphasize that the cross section \eq{numerator-a} to produce an additional gluon obviously vanishes in the absence of interaction with the nucleus, \ie, in the formal limit $S^{(i)} \to 1$. Thus,  \eq{numerator-a} could be equivalently expressed in terms of $T$-matrix elements ($T = S-1$) by formally replacing $S \to T$.

\subsection{Parton elastic scattering cross section}
\label{NoRadiation}

\begin{figure}
\centering
{\includegraphics[width=0.35\textwidth]{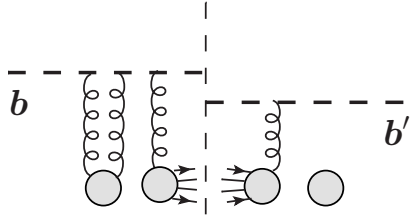}}
\centering
\caption{Amplitude squared for the elastic scattering of the energetic parton $a=q, g$ (represented by the dashed line) off a nucleus.}
\label{fig:Elastic}
\end{figure}

We will now express the denominator of \eq{spec-pA} corresponding to parton elastic scattering,
which amplitude squared is shown in Fig.~\ref{fig:Elastic}. In the quark case, the cross section for elastic scattering off a nucleus has been calculated in the saturation formalism in Ref.~\cite{Dumitru:2002qt}. In the notations of the previous section,\footnote{In particular, the quark transverse position is ${\bvec}$ in the amplitude and ${\bvec'}$ in its conjugate, and $\pvec_{0} = \zerovec$.} the quark-nucleus scattering cross section (convoluted with the quark distribution in the proton) is given by 
\be 
\label{no-rad}
\frac{d\sigma({\rm pA} \to q + X)}{dy \, d^2\pvec} = \frac{x_{\rm p} f_{q/{\rm p}}(x_{\rm p},\mu_{_{\rm F}}^2)}{(2\pi)^2} \, \int_{\bvec} \int_{\bvec'} \, e^{-i \pvec \cdot (\bvec - \bvec')} \frac{1}{N_c} \left< \tr \left[ \left(U_F(\bvec) - 1\right) (U_F^\dagger(\bvec') - 1 ) \right] \right>  \, .
\ee

Assuming a homogeneous target nucleus of {\it infinite} transverse extension, the correlator $\ave{\tr \left( U_F(\bvec) \right) }$ appearing in the integrand of \eq{no-rad} is independent of $\bvec$~\cite{Gelis:2001da}, and thus formally contributes to $[\delta^{(2)}(\pvec)]^2$ in \eq{no-rad}. In the large $p$ limit \eq{hard-limit}, all such terms can be dropped, leaving only the 2-point correlator $\propto \ave{ \tr ( U_F(\bvec) U_F^\dagger(\bvec') )}$. Clearly, gluon scattering can be treated in the same way, leading to the cross section for the elastic scattering of parton $a$, 
\be 
\label{denominator-a}
\frac{d\sigma({\rm pA} \to a + X)}{dy \, d^2\pvec} = \frac{x_{\rm p} f_{a/{\rm p}}(x_{\rm p},\mu_{_{\rm F}}^2)}{(2\pi)^2} \, \int_{\bvec} \int_{\bvec'} \, e^{-i \pvec \cdot (\bvec - \bvec')} \, {S}_a^{(2)}[{\bvec},{\bvec'}] \, .
\ee
The differential cross section for parton elastic scattering at large $\pvec$ is related to the Fourier transform of the 2-point correlator ${S}_a^{(2)}$.
 
In the large $\pvec$ limit \eq{hard-limit}, the size of the dipole $\rvec = {\bvec'} - {\bvec}$ in \eq{denominator-a} is small, $|\rvec| \lsim 1/p \ll 1/Q_s < 1/\lqcd$, and thus much smaller than any nuclear size $R_{\rm A}$, including that of the proton $R_{\rm p} \sim 1/\lqcd$. The above approximation of infinite transverse nuclear size is thus a good approximation, with finite size corrections expected to be at most 
$\sim {\cal O}({\lqcd^2}/{p^2})$. 

\subsection{Master equation}
\label{sec:master}

The induced soft radiation spectrum \eq{spec-pA} is obtained by taking the $z \ll 1$ limit of \eq{numerator-a}, dividing by \eq{denominator-a} and applying the medium-induced prescription \eq{ind-prescription}.

Within the approximation of an infinite transverse nuclear size (valid at large $p$), the two-point correlator $S^{(2)}_{a}[{\bvec},{\bvec'}]$ is a function of ${\bvec'}-{\bvec}$ only, $S^{(2)}_{a}[{\bvec},{\bvec'}] \equiv S_{a}({\bvec'}-{\bvec})$, where $S_a(\xvec)$ is given in appendix~\ref{app:Sformulae}, see \eq{dipole-Smatrix}. Similarly, the three-point correlator $S^{(3)}_{a}[{\bvec},{\xvec},\vvec']$ depends only on {\it relative} transverse positions, as shown by its explicit expression \eq{S3-a} (or equivalently \eq{S3-a-formal}). Changing variable ${\xvec} \to {\xvec} + \bvec'$ followed by $\bvec' \to \bvec + \rvec$ in \eq{numerator-a}, and using $z \Phi_a^g(z) = 4 C_R$ when $z \ll 1$ (see \eq{splittings}), the induced spectrum reads\footnote{A factor $\int d^2\bvec = S_\perp$, with $S_\perp$ the transverse area of the nuclear target, cancels out in the ratio between \eq{numerator-a} and \eq{denominator-a}.}$^{,}$\footnote{\label{foot:smallz}Note that when $z\ll 1$, one can safely replace $1-z \to 1$ in quantities such as $(1-z)\rvec$, $(1-z)\xvec$, etc. However, $z$ should obviously be kept in quantities like $z \xvec+\rvec$.}
\be 
\label{spec-ind-1}
\left. z \frac{dI}{dz} \right|_{\rm ind} = \left. \frac{2 C_R \alpha_s }{\pi^2} \, \frac{\int_{\rvec} e^{i \pvec \cdot \rvec} \!\int_{\xvec} \frac{{\xvec} \cdot ({\xvec} + \rvec)}{{\xvec}^2 ({\xvec} + \rvec)^2} \left\{S_{a}(\rvec) - S^{(3)}_{a}[\zerovec,\xvec+\rvec,z \xvec+\rvec] \right\}}{\int_{\rvec} \, e^{i \pvec \cdot \rvec } \, {S}_a(\rvec)} \right|_{\rm ind} \! \! \! \! .
\ee
The expression \eq{spec-ind-1} is the induced, soft ($z \ll 1$) radiation spectrum associated with the hard ($p \gg Q_s$) scattering of a parton $a=q, g$, as predicted in the saturation formalism. It is valid in the forward limit ($p^+ \to \infty$ at fixed $p \equiv |\pvec|$), up to finite size corrections of relative order $\sim {\cal O}({\lqcd^2}/{p^2})$.

The two terms in the bracket of \eq{spec-ind-1} are, respectively, the combined contribution from initial state and final state radiation (when $z \ll 1$ these contributions are identical), and the contribution from the interference. It will prove useful in the following section to rewrite \eq{spec-ind-1} and express the spectrum as the sum
\be
\label{spec-ind-Tmatrix}
\left. z \frac{dI}{dz} \right|_{\rm ind} = \left. z \frac{dI}{dz} \right|_{\rm IS}  +  \left. z \frac{dI}{dz} \right|_{\rm FS} +  \left. z \frac{dI}{dz} \right|_{\rm INT} \, ,
\ee
where the separate initial state (final state) and interference contributions, 
\bea
&& \hskip 0mm \left. z \frac{dI}{dz} \right|_{\rm IS} = \left. z \frac{dI}{dz} \right|_{\rm FS} = \frac{C_R \alpha_s}{\pi^2} \left. \frac{\int_{_\rvec} \, e^{i \pvec \cdot \rvec} \left[ S_a(\rvec)  -1 \right] \, \int_{_\xvec} \, \frac{{\xvec} \cdot ({\xvec} + \rvec)}{{\xvec}^2 ({\xvec} + \rvec)^2}}{\int_{_\rvec} \, e^{i \pvec \cdot \rvec} S_a(\rvec)} \right|_{\rm ind} \!\! ,  \label{spec-ind-IS} \\
&& \left. z \frac{dI}{dz} \right|_{\rm INT} = \frac{2C_R \alpha_s}{\pi^2} \left. \frac{\int_{_\rvec} \, e^{i \pvec \cdot \rvec} \int_{_\xvec} \, \frac{{\xvec} \cdot ({\xvec} + \rvec)}{{\xvec}^2 ({\xvec} + \rvec)^2} \left\{ 1 -  S^{(3)}_{a}[\zerovec,\xvec+\rvec,z \xvec+\rvec]  \right\} }{\int_{_\rvec} \, e^{i \pvec \cdot \rvec} S_a(\rvec)} \right|_{\rm ind} \! \!  , \hskip 1cm
\label{spec-ind-INT}
\eea
formally vanish when $S_a \to 1$, \ie\ in the absence of interaction with the target. 

\section{Explicit derivation of the induced soft radiation spectrum}
\label{sec:eval}

Here we present a brief derivation of \eq{spec-ind-IS} and \eq{spec-ind-INT} in the hard scattering limit \eq{hard-limit}. 
The calculation of sections~\ref{sec:ISFS} and~\ref{sec:int} is somewhat formal, but allows one to rapidly reach the result, which is confirmed by a more rigorous calculation presented in appendix~\ref{app:detailed}, where the order of magnitude of correction terms is also determined. The final result for the spectrum is summed up in section~\ref{sec:sumup}, see \eq{spec-master-pointlike}.

\subsection{Contribution from initial/final state radiation}
\label{sec:ISFS}

In the hard scattering limit \eq{hard-limit}, the contribution \eq{spec-ind-IS} from purely initial (final) state radiation can be evaluated as follows. (We recall the notation $p = |\pvec|$, $r= |\rvec|$, etc.)

The $\rvec$-integral in the numerator of \eq{spec-ind-IS} is dominated by $r \lsim 1/p \ll 1/Q_a$ (where $Q_a$ is the parton saturation scale in the nucleus, see footnote~\ref{foot:sat-def}), and we can thus approximate $S_a(\rvec)  -1 \propto Q_a^2 \, r^2$, see \eq{S-small-x}, up to corrections of relative order $\sim \morder{Q_a^2 \, r^2} \sim \morder{Q_a^2/p^2}$. Since the denominator of \eq{spec-ind-IS} is also $\propto Q_a^2$ (with the same accuracy, see \eq{Stilde-final-A}), the $Q_a$-dependence cancels in the ratio, leaving no contribution in the {\it induced} spectrum defined by \eq{ind-prescription}, up to terms $\sim \morder{Q_{a}^2/p^2}$.

The latter derivation is somewhat formal because \eq{spec-ind-IS} is actually ill-defined, the $\xvec$-integral  in the numerator being logarithmically divergent at large $\xvec$. A rigorous calculation, obtained by first regularizing this infrared divergence, and presented in appendix~\ref{app:ISFS}, confirms that the contribution to the induced spectrum from purely initial (final) state radiation is power-suppressed at large $p$, and allows one to estimate more accurately the magnitude of this contribution, namely (see \eq{IS-magnitude}), 
\be
\label{IS-magnitude-0}
\left. z \frac{dI}{dz} \right|_{\rm IS} = \left. z \frac{dI}{dz} \right|_{\rm FS} \sim \alpha_s \, \morder{\frac{{Q}_{a}^2 }{p^2} \log^2{\left( \frac{p^2}{\mu^2}\right)} }\, ,
\ee
where the parameter $\mu \sim \lqcd$ is the infrared regulator (see appendices \ref{app:x-space} and \ref{app:ISFS}). The initial (final) state contribution \eq{IS-magnitude-0} turns out to be subleading compared to the interference contribution derived in the next section. 

\subsection{Contribution from the interference term}
\label{sec:int}

The contribution \eq{spec-ind-INT} can be rewritten using \eq{S3-a-formal} (see however footnote~\ref{foot:formal}) as
\bea
\label{spec-ind-INT-1}
\left. z \frac{dI}{dz} \right|_{\rm INT} = \left. \frac{2 C_R \alpha_s }{\pi^2} \, \frac{\int_{\rvec} e^{i \pvec \cdot \rvec} \!\int_{\xvec} \frac{{\xvec} \cdot ({\xvec} + \rvec)}{{\xvec}^2 ({\xvec} + \rvec)^2} \left\{1 - \left[S_a({\xvec}+\rvec) S_a({\xvec})\right]^{\sigma} S_a(z{\xvec}+\rvec)^{\rho} \right\}}{\int_{\rvec} \, e^{i \pvec \cdot \rvec } \, {S}_a(\rvec)} \right|_{\rm ind} \! \! \! \! , \hskip 5mm  && \\
\label{rhosigdef}
\rho \equiv \frac{2C_R-C_A}{2 C_R} \ ; \hskip 10mm  \sigma \equiv 1-\rho = \frac{C_A}{2C_R} \, . \hskip 35mm  &&
\eea

At large $p$ the contribution \eq{spec-ind-INT-1} is dominated by $r \lsim 1/p \ll 1/Q_a$. When $x$ is of the same order, $x \sim r \lsim 1/p$, all dipole scattering factors appearing in \eq{spec-ind-INT-1} can be expanded at small values of their argument, leading to a contribution $\sim \morder{Q_a^2/p^4}$ in the numerator of \eq{spec-ind-INT-1}. Similarly to the case of initial state radiation discussed in the previous section, such a contribution cancels in the induced spectrum and leaves only terms $\sim \morder{Q_a^2/p^2}$ (up to logarithms). We infer that the contributions to \eq{spec-ind-INT-1} which are not power-suppressed must arise from an integration domain where $r \ll x$. 

Approximating the integrand of \eq{spec-ind-INT-1} in the latter domain we get 
\be
\label{spec-ind-INT-2}
\left. z \frac{dI}{dz} \right|_{\rm INT} \simeq 2C_R \, \frac{\alpha_s}{\pi^2} \left. \frac{\int_{_\rvec} \, e^{i \pvec \cdot \rvec} \int_{_\xvec} \, \frac{1}{\xvec^2} \left[ 1 -  S_a({\xvec})^{2\sigma} S_a(z{\xvec} + \rvec)^{\rho} \right]}{\int_{_\rvec} \, e^{i \pvec \cdot \rvec} S_a(\rvec)} \right|_{\rm ind} \, . 
\ee
In the numerator of \eq{spec-ind-INT-2}, the first term in the bracket gives a contribution $\propto \delta^{(2)}(\pvec)$ and can be dropped. Similarly, if $r$ were much smaller than $z x$, $S_a(z{\xvec} +\rvec)$ could be expanded at small $r$, yielding only analytic terms in $r$ and contributions of the type $\propto\int_{_\rvec} \, e^{i \pvec \cdot \rvec} \, r^{2n} = (2\pi)^2(-{\bf \nabla}_{\!\!\pvec}^2)^{^n} \delta^{(2)}(\pvec)$, which are irrelevant at large $p$. This suggests that the only way to obtain a potential contribution from \eq{spec-ind-INT-2} at large $p$ is to probe the {\it non-analytic} behavior in $r$ of $S(z{\xvec}+\rvec)$, which can be highlighted by changing variable $\rvec \to \rvec - z{\xvec}$ in the second term of \eq{spec-ind-INT-2}, leading to
\be
\label{spec-ind-INT-3}
\left. z \frac{dI}{dz} \right|_{\rm INT} \simeq - 2 C_R \, \frac{\alpha_s}{\pi^2} \left. \frac{\int_{_\rvec} \, e^{i \pvec \cdot \rvec} \, S_a(\rvec)^{\rho} \int_{_\xvec} \, \frac{S_a({\xvec})^{2\sigma}}{\xvec^2} \, e^{-i z \pvec \cdot \xvec}  }{\int_{_\rvec} \, e^{i \pvec \cdot \rvec} S_a(\rvec)} \right|_{\rm ind} \, . 
\ee
Expanding $S_a(\rvec)$ at small $r$ we easily find\footnote{We note that in the l.h.s. of \eq{ratio-rho}, both numerator and denominator are $\sim Q_a^2/p^4$ and not $\propto \delta^{(2)}(\pvec)$. This is because when $r \lsim 1/p \ll 1/Q_a$, the small-$r$ expansion of $S_a(\rvec)$ (obtained by expanding \eq{S-small-x}) is non-analytic in $r$, $S_a(\rvec) - 1 \propto Q_a^2 \, r^2 \log(\mu r)$. This crucial feature would be lost if performing the small-$r$ expansion at the level of Eq.~\eq{spec-ind-INT-2}.}
\be
\label{ratio-rho}
\frac{\int_{_\rvec} \, e^{i \pvec \cdot \rvec}  S_a(\rvec)^{\rho} }{\int_{_\rvec} \, e^{i \pvec \cdot \rvec}  S_a(\rvec)} \simeq \rho = \frac{2C_R-C_A}{2 C_R} \, , 
\ee
and finally using $S_a({\xvec})^{2\sigma} = S_a({\xvec})^{\frac{C_A}{C_R}} = S_g({\xvec})$, we arrive at
\be
\label{spec-master}
\left. z \frac{dI}{dz} \right|_{\rm INT} =  (2 C_R -C_A) \, \frac{\alpha_s}{\pi^2}  \int \frac{d^2\xvec}{\xvec^2} \, e^{-i z \pvec \cdot \xvec} \, \left[ S_{g {\rm p}}({\xvec}) - S_{g}({\xvec}) \right] \, .
\ee

Similarly to the previous section, the derivation presented here is quite formal (in particular, the contribution $\propto \delta^{(2)}(\pvec)$ neglected in \eq{spec-ind-INT-2} is actually multiplied by the divergent factor $\int_{_\xvec} \frac{1}{\xvec^2}$). However, the expression \eq{spec-master} also follows from the rigorous calculation presented in appendix~\ref{app:int} (see \eq{spec-ind-final-app}), up to terms which do not exceed the magnitude of the contribution from purely initial state radiation given by \eq{IS-magnitude-0}.

\subsection{Sum up}
\label{sec:sumup}

In summary, in the kinematical limit defined by \eq{hard-limit} and \eq{soft-limit}, the induced coherent spectrum is dominantly given by the interference contribution \eq{spec-master}. For the sake of simplicity, let us write the induced coherent spectrum associated with the scattering of parton $a$ off nucleus ${\rm A}$ with respect to a fictitious target obtained by setting $Q_{g {\rm p}} = 0$ in \eq{spec-master},\footnote{The spectrum \eq{spec-master} relevant to \pA\ collisions is then simply given by the difference of the spectrum \eq{spec-master-pointlike} between a nucleus and a proton target.} 
\be
\label{spec-master-pointlike}
z \frac{dI}{dz} =  (2 C_R -C_A) \, \frac{\alpha_s}{\pi^2}  \int \frac{d^2\xvec}{\xvec^2} \, e^{-i z \pvec \cdot \xvec} \, \left[ 1 - S_{g}({\xvec}) \right] \, .
\ee
The latter expression, with $S_{g}({\xvec})$ given by \eq{S-x}, coincides with the result found previously in Ref.~\cite{Peigne:2014uha} using the opacity expansion. 

The $\xvec$-integral in \eq{spec-master-pointlike}, which is a function of the three dimensionful parameters $\mu$, ${Q}_{g} \equiv Q_s$, and $z p$, was studied in Ref.~\cite{Peigne:2014uha}.\footnote{\label{foot:param}In Ref.~\cite{Peigne:2014uha}, the independent parameters were chosen to be $zp$, $\mu$, and the dimensionless parameter $L/\lambda_g = Q_s^2/(2\mu^2)$ appearing in the expression of $S_{g}({\xvec})$ (see \eq{S-x}), where $L$ is the path-length of the parton across the nucleus and $\lambda_g$ the elastic {\it gluon} mean free path.} Here we only quote the parametric limits of \eq{spec-master-pointlike} at fixed $Q_s^2/\mu^2 \gg 1$, at both `small' and `large' $z$ (still keeping $z \ll 1$), namely~\cite{Peigne:2014uha},
\begin{numcases}{z \frac{d I}{d z} \simeq (2C_R-C_A) \, \frac{\alpha_s}{\pi} \, \times}
\log{\left( \frac{{\widetilde Q}_s^2}{2 z^2 p^2 } \right)}   & \text{if $z \ll \frac{{\widetilde Q}_s}{p}$,} \label{case-1} \\
\hskip 6mm \frac{Q_s^2}{2 z^2 p^2 }  & \text{if $z \gg \frac{Q_s}{p}$,} \label{case-2}
\end{numcases}
where in \eq{case-1} we defined 
\be
\label{Qs-tilde}
{\widetilde Q}_s^2 \equiv Q_s^2 \log{\left(\frac{Q_s^2}{2\mu^2}\right)} \, .
\ee
Note that \eq{case-1} was derived in the logarithmic accuracy, \ie\ assuming logarithms (that in \eq{case-1} as well as in \eq{Qs-tilde}) to be much larger than unity. Thus, the difference between ${\widetilde Q}_s^2$ and $Q_s^2$ may seem irrelevant in the logarithm of \eq{case-1}, since $\log{\log{({Q_s^2}/{2\mu^2})}}$ becomes much larger than unity only for unrealistically large values of $Q_s$. However, keeping ${\widetilde Q}_s^2$ in \eq{case-1}, in addition to being parametrically correct in the academic limit where $\log{\log{({Q_s^2}/{2\mu^2})}} \gg 1$, turns out to be numerically relevant, since it gives a more accurate small-$z$ approximation to the exact expression \eq{spec-master-pointlike} even for realistic values of the parameters~\cite{Peigne:2014uha}.

We stress that the order of corrections to the result \eq{spec-master-pointlike} was not estimated in Ref.~\cite{Peigne:2014uha}. As mentioned in the previous sections and shown in appendix~\ref{app:detailed}, those corrections have a (maximal) magnitude given in \eq{IS-magnitude-0}, and are thus power-suppressed at large $p$. Therefore, the spectrum \eq{spec-master-pointlike} is valid as long as its (power-suppressed) large-$z$ behavior \eq{case-2}  dominates over \eq{IS-magnitude-0}, \ie\ as long as $z < 1/\log{({p^2}/{\mu^2})}$. The window where \eq{case-2} is valid thus reads
\be
\label{window}
\frac{Q_s}{p} < z <  \frac{1}{\log{( \frac{p^2}{\mu^2})}} \, ,
\ee
and actually exists provided $p > Q_s \log{({p^2}/{\mu^2})}$. Note that when the latter condition is not satisfied (but still $p \gg Q_s$), the validity of the spectrum \eq{spec-master-pointlike} is limited to its small-$z$ logarithmic behavior \eq{case-1}. Indeed, at small enough $z$ the latter always dominates over correction terms such as \eq{IS-magnitude-0}, which remain bounded at small $z$. 

\section{Comparing the opacity expansion and saturation formalisms}
\label{sec:comp-form}

Let us emphasize that the calculation of the induced coherent spectrum \eq{spec-master-pointlike} may look quite different in the saturation formalism (used in the present study) and in the opacity expansion (used in Ref.~\cite{Peigne:2014uha}). In the saturation formalism, the calculation is performed in coordinate space and scatterings are resummed (`exponentiated') from the start in the two-point and three-point correlators entering \eq{spec-ind-1}. In contrast, the calculation of Ref.~\cite{Peigne:2014uha} is done in momentum space, for a single {\it hard} scattering and $n$ {\it soft} rescatterings. The calculation at first order in opacity ($n=1$) is generalized to any order $n$ by making use of recurrency relations. Summing over $n$ then leads to the usual exponentiation (producing the term $S_{g}({\xvec})$ in \eq{spec-master-pointlike}), which exponentiation thus appears at a later stage of the calculation when compared to the saturation formalism. 
 
In order to make the equivalence between formalisms more transparent, we show below how the coherent spectrum derived in~\cite{Peigne:2014uha} is linked to the expression \eq{spec-ind-1} in the saturation formalism. Since we have shown in section~\ref{sec:eval} that the induced spectrum arises dominantly from the interference term, we may start from \eq{spec-ind-INT}, and insert the expressions \eq{dipole-Smatrix} and \eq{S3-a} of the correlators. As shown in section~\ref{sec:int}, at large $p$ the two-point correlator in the denominator of \eq{spec-ind-INT} can be formally expanded to first order in the function $\hat \Gamma$ (defined in \eq{Gamma-2}). Expanding then the numerator of \eq{spec-ind-INT} in $\hat \Gamma$, the linear term gives no contribution to the {\it induced} spectrum (for the reason mentioned in section~\ref{sec:int}, see the discussion after \eq{rhosigdef}), so that we keep only terms of power $m \geq 2$ in $\hat \Gamma$, corresponding to the contribution of $m$ scatterings (associated with the symmetry factor $1/m!$). We thus rewrite \eq{spec-ind-INT} as (note that the `induced prescription' with respect to a fictitious target having $Q_{s} = 0$ is now irrelevant) 
\be
\label{eq:comp-form-1}
z \frac{dI}{dz} =  \mathop{\sum}_{m \geq 2} \frac{2 \alpha_s }{\pi^2} \, \frac{\int_{\rvec} e^{i \pvec \cdot \rvec} \!\int_{\xvec} \frac{{\xvec} \cdot ({\xvec} + \rvec)}{{\xvec}^2 ({\xvec} + \rvec)^2} \frac{(-1)^m}{m!} \! \left\{ \frac{C_A}{2} \, [\hat{\Gamma}(\xvec) + \hat{\Gamma}(\xvec +\rvec) ] + \frac{2 C_R - C_A}{2} \, \hat{\Gamma}(z{\xvec}+\rvec) \right\}^m}{\int_{\rvec} \, e^{i \pvec \cdot \rvec } \, \hat \Gamma(\rvec)} \, .
\ee

As in section~\ref{sec:int}, we change variable $\rvec \to \rvec - z{\xvec}$ in the numerator, and note that at large $p$ the dominant contribution to \eq{eq:comp-form-1} arises from the domain $r \ll x$, leading to
\be
\label{eq:comp-form-2}
z \frac{dI}{dz} = \mathop{\sum}_{m \geq 2} \frac{2\alpha_s }{\pi^2} \, \frac{\int_{\rvec} e^{i \pvec \cdot \rvec} \!\int \frac{d^2\xvec}{\xvec^2} \, e^{-i z \pvec \cdot \xvec} \, \frac{(-1)^m}{m!} \! \left\{  \frac{2 C_R - C_A}{2} \, \hat{\Gamma}(\rvec) + C_A \, \hat{\Gamma}(\xvec) \right\}^m}{\int_{\rvec} \, e^{i \pvec \cdot \rvec } \, \hat \Gamma(\rvec)}  \, .
\ee
When expanding the bracket of \eq{eq:comp-form-2}, the term $\propto \hat{\Gamma}(\xvec)^m$ formally contributes to $\delta^{(2)}(\pvec)$ which can be dropped at large $p$, and the terms $\propto {m \choose k}  \, \hat{\Gamma}(\xvec)^{m-k} \, \hat{\Gamma}(\rvec)^k$ (for $1 \leq k \leq m$) scale as $\morder{(\bar{Q}^2/p^2)^k}$ (recall that $r \sim 1/p$). Thus, the dominant term is given by the term $\propto m \, \hat{\Gamma}(\xvec)^{m-1} \, \hat{\Gamma}(\rvec)$, which has a simple physical interpretation. The factor $\hat{\Gamma}(\rvec)$ singles out the {\it hard} scattering (with the factor ${m \choose 1} = m$ accounting for the $m$ ways to choose one scattering among $m$), and the factor $\hat{\Gamma}(\xvec)^{m-1}$ corresponds to the additional $n=m-1$ {\it soft} rescatterings. This contribution should thus coincide with the order $n$ of the opacity expansion used in Ref.~\cite{Peigne:2014uha}. We also remark that the hard scattering factor $\int_{\rvec} e^{i \pvec \cdot \rvec} \, \hat{\Gamma}(\rvec)$ then cancels between numerator and denominator in \eq{eq:comp-form-2}, \ie\ between the radiative and elastic cross sections (see also our comments after \eq{Stilde-final-A}), leading to 
\be
\label{eq:comp-form-3}
z \frac{dI}{dz} = (2 C_R - C_A) \,  \frac{\alpha_s }{\pi^2}  \int \frac{d^2\xvec}{\xvec^2} \, e^{-i z \pvec \cdot \xvec} \,\mathop{\sum}_{n \geq 1}  \frac{(-1)^{n+1}}{n!} \left[ C_A \, \hat{\Gamma}(\xvec) \right]^n  \, , 
\ee
which using \eq{dipole-Smatrix} directly reproduces the result \eq{spec-master-pointlike}. 

Thus, the above provides a simple derivation of the induced coherent spectrum, where the equivalence between the opacity expansion and saturation formalisms is explicit. For instance, the term of first order in opacity ($n=1$) in \eq{eq:comp-form-3} can be rewritten using \eq{Gamma-2} and $C_A {\bar Q}^2/(2\mu^2) = L/\lambda_g$ (see footnote~\ref{foot:param}) as
\be
\label{eq:comp-form-4}
\left. z \frac{dI}{dz} \right|_{n=1} =  (2 C_R - C_A) \, \frac{\alpha_s }{\pi^2}  \, \frac{L}{\lambda_g} \int d^2 \ellvec \, V(\ellvec)  \!\int \frac{d^2\xvec}{\xvec^2} \, e^{-i z \pvec \cdot \xvec} \, \left(1-e^{i {\ellvec} \cdot {\xvec}} \right)  \, ,
\ee
where we defined $V(\ellvec) \equiv \frac{\mu^2}{\pi (\ellvec^2 + \mu^2)^2}$. Shifting now from the transverse coordinate space (of the radiated gluon) to momentum space by making use of the identity (see \eq{conformal-identity}) 
\be
\frac{1}{\xvec^2} = \frac{\xvec}{\xvec^2} \cdot \frac{\xvec}{\xvec^2} = \int \frac{d^2 \kvec}{2i \pi} \int \frac{d^2 \kvec'}{2i \pi} \, \frac{\kvec}{\kvec^2} \cdot  \frac{\kvec'}{\kvec'^2} \, \,  e^{i (\kvec + \kvec') \cdot \xvec} \, ,
\ee
we easily obtain
\be
\label{spectrum-n1}
\left. z \frac{dI}{dz}  \right|_{n=1} =  (2 C_R - C_A) \, \frac{\alpha_s}{\pi^2} \, \frac{L}{\lambda_g} \int d^2 \ellvec \, V(\ellvec) \int d^2 \kvec \, \left[ \frac{\kvec -  \ellvec}{(\kvec - \ellvec)^{2}} -  \frac{\kvec}{\kvec^{2}} \right] \cdot \frac{-(\kvec - z\pvec)}{(\kvec - z\pvec)^{2}} \, ,
\ee
which corresponds exactly to the momentum-space expression of the induced coherent spectrum at first order in opacity in Ref.~\cite{Peigne:2014uha}.\footnote{See Eq.~(29) of Ref.~\cite{Peigne:2014uha}.}

Finally, we stress that it is the chosen kinematical limit \eq{hard-limit} which leads to the dominance of a {\it single} hard exchange (in both the radiative and elastic cross sections), the other scatterings being soft, independently of the formalism used. The probability to have more than one hard scattering is suppressed by $\morder{Q_{s}^2/p^2}$ (at least), which specifies the order of magnitude of the correction terms to the spectrum \eq{spec-master-pointlike}, as shown in section~\ref{sec:eval} and appendix~\ref{app:detailed}. 

\section{Outlook: implementing small-$x$ evolution}
\label{sec:smallx}

In this paper, we treated the scattering of the quarks and gluons off the target nucleus in the approximation where the exchanged gluons couple eikonally to the incoming partons. Technically, these exchanges are encoded in products of Wilson lines averaged over the fluctuations of the field of the nucleus, see Eqs.~\eq{S4-corr}-\eq{S3-corr-2}. To arrive at explicit formulae, we used the Gaussian approximation for the field of the target, allowing one to express all correlators of Wilson lines with the help of {\it two-point} correlators having the simple McLerran-Venugopalan (MV) form~\cite{McLerran:1993ni}, see appendix~\ref{app:Sformulae}. As is well-known, those correlators depend on a saturation scale $Q_s$ (with $Q_s^2 = Q_g^2 = C_A \, {\bar Q}^2$ in our convention, see \eq{Qbar-def}), which in the MV model is independent of the parton-nucleus collision energy.   

In principle, the dependence on the collision energy (or {\it energy evolution}), could be readily implemented for a process such as medium-induced gluon radiation. In the nucleus rest frame and at high collision energy, the incoming parton radiates not only one but many soft gluons. One may attribute the induced energy loss to the hardest gluon, and factorize the softer gluons (associated with shorter lifetimes) into the nuclear parton density, which then captures the energy evolution (also named small-$x$ evolution). How this comes about technically was shown in detail in the case of the broadening process (in the large-$N_c$ limit) in Ref.~\cite{Mueller:2012bn}, where it was argued that the same should occur for similar semi-inclusive processes at leading and next-to-leading logarithmic accuracy.

Let us describe how small-$x$ evolution could be implemented for the induced energy loss process. To this aim, imagine we start with a scattering at small rapidity. In this low-energy regime, it is legitimate to represent the field of the nucleus by a Gaussian field, as in the MV model. Increasing the collision energy by boosting the nucleus leads to a modification of the nucleus field driven by the B-JIMWLK equation. Hence the correlators appearing in the equations should be taken, at high energy, as solutions of the B-JIMWLK equations. Note that they must be evaluated numerically since no analytical solution to the latter equations is known. It is possible to simplify the formalism by going to the large-$N_c$ limit, where all correlators reduce to (sums of products of) two-point functions being solutions of the simpler Balitsky-Kovchegov equation, which however still needs numerical evaluation. Moreover, let us stress that in the large-$N_c$ limit, the induced coherent energy loss \eq{spec-master-pointlike} (derived in the Gaussian approximation) would vanish in the {\it quark} case (since $2 C_F -C_A = -1/N_c \to 0$ when $N_c \to \infty$). Thus, the large-$N_c$ limit is too drastic to address interesting effects such as a {\it negative} induced radiation spectrum associated with the $q \to q$ process, as found in~\cite{Peigne:2014uha} and confirmed by the present study. 

In order to address the effect of small-$x$ evolution on induced energy loss (without relying on the large-$N_c$ limit), one may still assume a Gaussian approximation for the solution of the evolution itself, which was shown \cite{Iancu:2011nj} to be practically a good approximation. Then, it is enough to use the MV form for the two-point functions, but with a saturation scale now promoted to an $x$-dependent function.

The value of $x$ that should be used for the evolution of the B-JIMWLK correlators (or as an argument of the saturation scale in the case of the Gaussian approximation) is the Bjorken-$x$ associated to a gluon 
of energy $p^{\prime+}$ and typical size ${\xvec}$ (namely the transverse size vector appearing in the integration~\eq{spec-master-pointlike}) scattering off a proton (of mass $m_{\rm p}$) of the nucleus, namely $x_{\rm B}=1/( 2m_{\rm p} \, {\xvec}^2 \, p^{\prime+})$~\cite{Liou:2014rha}.

In summary, we expect the radiation responsible for QCD small-$x$ evolution to be derived independently of the harder (though still soft compared to the radiating parton) induced coherent radiation studied in the present paper. However, how this works in practice remains to be studied. In particular, it is not known for sure to which extent the modification of the induced spectrum arising from small-$x$ evolution would be correctly captured by replacing the saturation scale $Q_s$ appearing in \eq{spec-master-pointlike} by an $x$-dependent scale $Q_s(x)$. Based on previous studies of other observables, we may expect the latter recipe to be valid at least in the small-$z$ region, where the parametric behaviour \eq{case-1} of the spectrum is logarithmic. Numerical studies using an implementation of the B-JIMWLK equation (see \eg\ Ref.~\cite{Rummukainen:2003ns}) could also help answering such questions. 

\acknowledgments

We thank Al Mueller and Heribert Weigert for useful discussions and comments. Feynman diagrams have been drawn with the JaxoDraw software~\cite{Binosi:2008ig}. This work was done in the framework of the `D\'efi InPhyNiTi' project GLUCOLPA. EP acknowledges support from: European Research Council grant HotLHC ERC-2011-StG-279579; Ministerio de Ciencia e Innovacion of Spain under project FPA2014-58293-C2-1-P; Xunta de Galicia (Conselleria de Educacion) within the Strategic Unit AGRUP2015/11.

\appendix

\section{Two-point and three-point correlators}
\label{app:Sformulae}

In this appendix we consider the generic case of a parton $a$ of color charge $C_R$ (with in practice $a=q$, $C_R=C_F$ or $a=g$, $C_R=C_A$), and give explicit expressions of the $S$-matrix elements for the scattering of an $a \bar a$ dipole and of a color-neutral $ag\bar a$ system off the nucleus. Throughout this appendix, we assume that the color sources in the nucleus have Gaussian correlations, which should be valid for asymptotically large nuclei (namely $A \gg 1$). We also assume a uniform distribution of the nuclear matter in the transverse plane.

In section~\ref{app:x-space} we quote the coordinate-space expressions of the two-point and three-point correlators known from previous studies, which also fixes our notations and conventions. In section~\ref{app:Stilde} we derive the Fourier transform of the two-point correlator in the large-momentum limit. 

\subsection{Coordinate-space expressions}
\label{app:x-space}

\subsubsection*{two-point correlator}

Let us start with the $a \bar a$ dipole scattering $S$-matrix defined in \eq{S2-corr}, namely (the normalized trace of) the correlator of two Wilson lines separated by the transverse distance $\xvec$. In the framework of the Gaussian approximation for the field of the nucleus, it takes the well-known McLerran-Venugopalan (MV) form~\cite{McLerran:1993ni}. Following appendix A.5.1 of Ref.~\cite{Blaizot:2004wv} we write the $S$-matrix element of a dipole of transverse size ${\xvec}$ as
\bea
\label{dipole-Smatrix}
&& \hskip 8mm S_a ({\xvec}) = e^{- C_R {\hat \Gamma} ({\xvec})} \, , \\ \nn \\
&& \hat \Gamma({\xvec}) \equiv {\bar Q}^2 \int \frac{d^2 \ellvec}{2\pi} \, \frac{1-e^{i\ellvec \cdot \xvec}}{(\ellvec^2)^{^2}} \, , \label{Gamma-1}
\eea
where we define  
\be
\label{Qbar}
{\bar Q}^2 \equiv \frac{g^4}{2\pi} \int_{-\infty}^\infty  \mu_{\rm A}^2(z^+) \, dz^+  \, ,
\ee
with $\mu_{\rm A}^2(z^+)$ the density per `unit volume' $dV \equiv d^2\rvec\, dz^+$ of color sources of the target nucleus A. The `saturation scale' of parton $a$ in the nucleus is simply related to the scale ${\bar Q}$ as
\be
\label{Qbar-def}
Q_a^2 \equiv C_R \, {\bar Q}^2 \, .
\ee

The function \eq{Gamma-1} has a logarithmic infrared divergence, but as is well-known, this divergence is effectively screened by color neutrality at distance scales larger than the nucleon size $\sim {\rm \Lambda}_{_{\rm QCD}}^{-1}$ (see for instance the discussion in appendix A of Ref.~\cite{Gelis:2001da}). We choose to regularize the divergence by introducing a `gluon mass' $\mu \sim {\rm \Lambda}_{_{\rm QCD}}$, namely by replacing ${\ellvec}^2 \to {\ellvec}^2 + \mu^2$, in the denominator of the propagator appearing in \eq{Gamma-1}. The latter thus becomes 
\be
\label{Gamma-2}
\hat \Gamma({\xvec}) = {\bar Q}^2  \int \frac{d^2 \ellvec}{2\pi} \frac{1-e^{i {\ellvec} \cdot {\xvec}}}{({\ellvec}^2 + \mu^2)^{^2}} =  \frac{{\bar Q}^2}{2 \mu^2} \left[ 1 - \mu x \, {\rm K}_1(\mu x) \right] \, . 
\ee
Within this regularization scheme, the $S$-matrix element \eq{dipole-Smatrix} reads
\be
\label{S-x}
S_a ({\xvec}) = \exp{\left\{ - \frac{Q_a^2}{2 \mu^2} \left[  1 - \mu x \, {\rm K}_1(\mu x)  \right] \right\}} \, .
\ee

It is useful to quote the behavior of $S_a({\xvec})$ at small $x \ll 1/\mu$. Using
\be
1 - \mu x \, {\rm K}_1(\mu x) \mathop{\simeq}_{x \ll 1/\mu} \left[ 1 + 2 \log{\left( \frac{C}{\mu x} \right)}\right] \frac{\mu^2 x^2}{4} + \morder{\mu^4 x^4  \log{\mu x}} \, ,
\ee 
where $C= 2\,e^{-\gamma}$ (with $\gamma$ the Euler constant), we obtain
\be
\label{S-small-x}
S_a({\xvec}) \mathop{\simeq}_{x \ll 1/\mu}  \exp\left[-\frac{Q_a^2 }{8} \, x^2 \log \left(\frac{1}{x^2 \mu^2} \right)\right] \, ,
\ee 
which coincides with the expression of the dipole scattering $S$-matrix in the MV model~\cite{McLerran:1993ni}.

\subsubsection*{three-point correlator}

Let us now turn to the $ag\bar a$ system (in practice $qg\bar q$ or $ggg$). In the Gaussian approximation, the three-point correlator $S^{(3)}_{a}[{\bvec},{\xvec},{\vvec'}]$ defined in \eq{S3-corr-1} (where $\bvec$, $\vvec'$ and $\xvec$ are the transverse coordinates of $a$, $\bar a$ and the gluon, respectively) may be expressed with the help of the function $\hat \Gamma$. The relevant relations can be found in 
Ref.~\cite{hep-ph/0106240} (see also Ref.~\cite{Marquet:2010cf}, and Ref.~\cite{Blaizot:2004wv} for the particular $qg\bar q$ case). We may cast the results as (with $a=q,g$)
\be
S^{(3)}_{a}[{\bvec},{\xvec},{\vvec'}] = \exp{\left\{-\frac{C_A}{2} \, {\hat \Gamma} (\bvec - \xvec) -\frac{C_A}{2} \, {\hat \Gamma} (\xvec - \vvec') -\frac{2C_R-C_A}{2} \, {\hat \Gamma} (\vvec' - \bvec) \right\} } \, ,
\label{S3-a}
\ee
which holds for any `generalized parton' $a$ in color representation $R$, as shown in Ref.~\cite{Marquet:2010cf}. 

Let us remark that using \eq{dipole-Smatrix}, the expression \eq{S3-a} may be rewritten as\footnote{\label{foot:formal}The expression \eq{S3-a-formal} of the three-point correlator in terms to the dipole $S$-matrix element \eq{dipole-Smatrix} simply arises from the fact that the color factors in \eq{S3-a} can always be put in the exponent of an exponential factor, but this is only formal and \eq{S3-a-formal} has no special physical interpretation. In particular, we stress that in the quark case ($a=q$), the third term in the exponential of \eq{S3-a} is positive (since $2C_F-C_A = -1/N_c$), and therefore the third factor in the r.h.s. of \eq{S3-a-formal} cannot be a true $S$-matrix element.}
\be
S^{(3)}_{a}[{\bvec},{\xvec},{\vvec'}] = \left[S_a({\bvec}-{\xvec})\right]^{\frac{C_A}{2C_R}} \left[S_a({\xvec}-{\vvec'})\right]^{\frac{C_A}{2C_R}} \left[S_a({\vvec'}-\bvec)\right]^{\frac{2 C_R -C_A}{2C_R}} \, .
\label{S3-a-formal}
\ee

\subsection{Dipole $S$-matrix element in the large-momentum limit}
\label{app:Stilde}

Here we study the large $p\equiv|\pvec|$ asymptotics of the Fourier transform ${\tilde S}_a(\pvec)$ of the two-point correlator $S_a(\rvec)$ defined by 
\be
\label{Stilde-A}
{\tilde S}_a(\pvec) = \int d^2 \rvec \, e^{i {\pvec} \cdot \rvec} S_a(\rvec) \equiv \int_{\rvec} \, e^{i {\pvec} \cdot \rvec} S_a(\rvec)  \, ,
\ee
where $S_a(\rvec)$ is given by \eq{S-x}. Note that the integral in \eq{Stilde-A} is perfectly convergent and well-defined. 

When $p$ is large, we expect that only small values of $r$, namely $r \lesssim 1/p \ll 1/\mu$, contribute to the integral. In order to compute the large-$p$ asymptotics, we may try and expand $S_a(\rvec)$ given by \eq{S-small-x} as a power series\footnote{The series~\eq{series-adiab-A} depends on the integrals $H_n$, which are actually ill-defined. A mathematically rigo\-rous procedure could consist, for instance, in introducing an exponential cutoff of the form $e^{-\varepsilon r}$ in the integral in \eq{Stilde-A} before expanding. Then all terms in the expansion would be well-defined. However, such complications can be avoided by using analyticity arguments such as the ones used in our derivation of $H_n$ below.}
\bea
\label{series-adiab-A}
&& {\tilde S}_a(\pvec) =   \sum_{n=0}^{\infty} \,  \frac{1}{n!} \left(- \frac{1}{4} \, Q_a^2 \, {\bf \nabla}_\pvec^2 \right)^n H_n \, , \\ 
&&H_n \equiv \int_{\rvec} \, e^{i \pvec \cdot \rvec} \,  \log^n(\mu r) \, .
\label{Hn-0}
\eea

The $n=0$ term in the expression \eq{series-adiab-A} gives a contribution $\propto \delta^{(2)}(\pvec)$, which can be dropped at finite $p$. The other terms can be evaluated as follows. In \eq{Hn-0} we use the representation 
\be
\label{rep-log-A}
\log^n(\mu r) = \left. \frac{\dd^n}{\dd \delta^n} \right|_{\delta = 0} \mu^{\delta} r^{\delta} \, ,
\ee
integrate over the angle between $\pvec$ and $\rvec$, and perform the change of variable $u=pr$ in the remaining integral over $r$, to obtain 
\be
\label{Hn}
H_n = \frac{2 \pi}{p^2} \,  \left. \frac{\dd^n}{\dd \delta^n} \right|_{\delta = 0} \left(\frac{\mu}{p}\right)^{\delta} \! \int_0^{\infty} \!  du \, u^{1+\delta} \, {\rm J}_0(u)  \, .
\ee
The integral over $u$ can be calculated for values of $\delta$ for which it is convergent, and then analytically continued in the vicinity of $\delta=0$. We arrive at
\be
\label{Hn-Fn}
H_n =  \frac{2 \pi}{p^2} \left. \frac{\dd^n}{\dd \delta^n} \right|_{\delta = 0} \left(\frac{\mu}{p}\right)^{\delta} \frac{2^{\delta} \, \delta \,\Gamma\left(\frac{\delta}{2}\right)}{\Gamma\left(-\frac{\delta}{2}\right)} = \frac{2 \pi}{p^2} \,
F^{(n)}(0) \, ,
\ee
expressed in terms of the $n^{\rm th}$ derivative in $\delta =0$ of the function $F(\delta)$ defined by 
\be
F(\delta) = e^{\delta A} \, f(\delta) \ ; \ \ \  A = \log{\left(\frac{2 \mu}{p}\right)} - \gamma \ ; \ \ \ f(\delta) = \frac{\delta \,\Gamma\left(\frac{\delta}{2}\right) e^{\gamma \, \delta}}{\Gamma\left(-\frac{\delta}{2}\right)}  \, .
\ee

The derivatives $F^{(n)}(0)$ are related to the derivatives $f^{(k)}(0)$ using the identity 
\be
\label{der-bin}
F^{(n)}(0) = \sum_{k=0}^{n} {{n}\choose{k}} A^{n-k} f^{(k)}(0) \, .
\ee
The derivatives $f^{(k)}(0)$ can be simply obtained by Taylor expanding $f(\delta)$ around $\delta=0$ using the identity (valid for $|\delta| < 1$)
\be
\label{f-simple}
f(\delta) \equiv \frac{\delta \,\Gamma\left(\frac{\delta}{2}\right) e^{\gamma \, \delta}}{\Gamma\left(-\frac{\delta}{2}\right)} 
= - \delta \exp{\left\{ -  \sum_{\ell=1}^{\infty} \frac{\zeta{(2\ell +1)}}{(2\ell +1) 4^\ell}  \, \delta^{2\ell +1} \right\}} \, ,
\ee
which can be derived from the Taylor expansion of the function $\log\Gamma(1+z)$ around $z=0$,
\be
\log{\Gamma(1+z)} = - \gamma z + \sum_{k=2}^{\infty} \frac{\zeta{(k)}}{k} (-z)^k \, ,
\ee
and using $\Gamma(1+z) =z\Gamma(z)$. Using \eq{f-simple} one finds 
\be
f^{(k)}(0)=\left\{ 0, -1, 0, 0, 2\zeta{(3)}, 0 \right\} {\rm \ \ for \ } k=0, \ldots, 5 \, .
\ee
Inserting this in \eq{der-bin} we obtain 
\be
F^{(n)}(0)=\left\{-1, -2A, -3A^2, -4A^3 + 2\zeta{(3)}, -5A^4 + 10 \zeta{(3)} A \right\} {\rm \ \ for \ } n=1, \ldots, 5 \, .
\ee
Using \eq{Hn-Fn} we get
\be
H_1 =  - \frac{2 \pi}{p^2} \ ; \ \  H_2 = \frac{4 \pi}{p^2} \left[ \log{\left(\frac{p}{2 \mu} \right)} + \gamma  \right]  \ ; \ \ 
H_3 = - \frac{6 \pi}{p^2} \left[ \log{\left(\frac{p}{2 \mu} \right)} + \gamma \right]^2 \, .
\label{H123}
\ee
Note that for $n \geq 4$, $H_n$ involves the $\zeta$-function of odd integers, for instance:
\be
H_4 =  \frac{8 \pi}{p^2} \left\{ \left[ \log{\left(\frac{p}{2 \mu} \right)} + \gamma \right]^3 + \frac{\zeta{(3)}}{2} \right\} \, .
\ee
Using \eq{H123} in \eq{series-adiab-A} and the identity ${\bf \nabla}_\pvec^2 \, f(p^2) =  \frac{\dd}{\dd p^2} \! \left(4p^2 \frac{\dd f(p^2)}{\dd p^2}\right)$, one finds the large $p$ behavior of ${\tilde S}_a(\pvec)$,
\be
\label{Stilde-final-A}
p \gg Q_a \ \ \ \Rightarrow \ \ \ {\tilde S}_a(\pvec) \simeq  \frac{2 \pi Q_a^2}{p^4} \left[ 1 +\morder{\frac{Q_a^2 }{p^2} \log{\left( \frac{p^2}{\mu^2}\right)}}\right] \, .
\ee

Note that the dominant term $\sim 1/p^4$ in \eq{Stilde-final-A} reflects the fact that at large $p$, the parton elastic scattering cross section \eq{denominator-a} is dominated by a {\it single} hard exchange. The associated factor $Q_a^2 \propto L$ can be interpreted as (proportional to) the probability to find, on the path length $L$, a scattering center where such a hard exchange occurs.

For further use we define
\be
\label{In-A}
I_n = \int_{\rvec} \, e^{i \pvec \cdot \rvec} \, r^2  \log^n(\mu r) = - {\bf \nabla}_\pvec^2 \, H_n \, ,
\ee
and quote the values of $I_1$, $I_2$, $I_3$ obtained using \eq{H123}:
\bea
I_1 &=& \frac{8 \pi}{p^4} \, , \label{I1-A}  \\
I_2 &=& - \frac{16 \pi}{p^4} \left[  \log{\left( \frac{p}{2 \mu} \right)} +\gamma -1 \right]  \, , \label{I2-A}  \\
I_3 &=& \frac{24 \pi}{p^4} \left[  \left( \log{\left( \frac{p}{2 \mu} \right)} +\gamma -1 \right)^2 - \frac{1}{2} \right] \, . \label{I3-A}
\eea

\section{Detailed derivation of the induced radiation spectrum}
\label{app:detailed}

\subsection{Contribution from purely initial/final state radiation}
\label{app:ISFS}

Here we evaluate the contribution \eq{spec-ind-IS} of purely initial (final) state radiation to the induced spectrum.

As discussed in section~\ref{sec:ISFS}, the integral $\int_{_\xvec} \, \frac{{\xvec} \cdot ({\xvec} + \rvec)}{{\xvec}^2 ({\xvec} + \rvec)^2}$ is divergent and needs to be re\-gularized. First, we note that using the identity
\be
\label{conformal-identity}
\frac{\xvec}{\xvec^2} = \int \frac{d^2 \kvec}{2i \pi} \, \frac{\kvec}{\kvec^2} \, e^{i \kvec \cdot \xvec} \, ,
\ee
the $\xvec$-integral can be formally rewritten as a $\kvec$-integral,
\be
\label{div-int}
\int d^2 \xvec \, \frac{{\xvec} \cdot ({\xvec} + \rvec)}{{\xvec}^2 ({\xvec} + \rvec)^2} = \int \frac{d^2 \kvec}{\kvec^2} \, e^{-i \kvec \cdot \rvec} \, ,
\ee
which can be regularized using the same parameter $\mu$ as that used in appendix~\ref{app:x-space}, namely, $\kvec^2 \to \kvec^2 + \mu^2$ in the r.h.s. of \eq{div-int}. Using
\be
\int \frac{d^2 \kvec}{\kvec^2 + \mu^2} \, e^{-i \kvec \cdot \rvec} = 2\pi \, {\rm K}_0(\mu r) \, ,
\ee
we thus rewrite \eq{spec-ind-IS} in the regularized form 
\be
\label{spec-ind-IS-reg}
\left. z \frac{dI}{dz} \right|_{\rm IS} = 2\frac{\alpha_s C_R}{\pi} \left. \frac{\int_{_\rvec} \, e^{i \pvec \cdot \rvec} \left[S_a (\rvec)  -1 \right]  \, {\rm K}_0(\mu r) }{\int_{_\rvec} \, e^{i \pvec \cdot \rvec}  S_a(\rvec)} \right|_{\rm ind} \, .
\ee

At large $\pvec$, the behavior of the denominator of \eq{spec-ind-IS-reg} was obtained in appendix~\ref{app:Stilde} by expanding $S_a(\rvec)$ at small $\rvec$, see \eq{Stilde-final-A}. Similarly, the large $\pvec$ limit of the numerator of \eq{spec-ind-IS-reg} can be found by expanding both $S_a(\rvec)$ and ${\rm K}_0(\mu r)$ at small $\rvec$. Expanding \eq{S-small-x} and using ${\rm K}_0(\mu r) = \log{\left( \frac{C}{\mu r} \right)} + \morder{\mu^2 r^2  \log{\left( \frac{1}{\mu r} \right)}}$  when $r \ll 1/\mu$ (with $C= 2\,e^{-\gamma}$), we obtain
\bea
\int_{_\rvec} \, e^{i \pvec \cdot \rvec} \left[S_a (\rvec)  -1\right] \, {\rm K}_0(\mu r)  &\simeq& \frac{Q_a^2}{4} \! \! \int_{_\rvec} \! e^{i \pvec \cdot \rvec} r^2 \left[ - \log^2(\mu r) + \log{C} \log(\mu r)  +  \morder{r^2 Q_a^2  \log^3(\mu r)}  \right]  \nn \\
&=&  \frac{2 \pi Q_a^2}{|\pvec|^4} \left[ \log{\left( \frac{p^2}{C e^2 \mu^2} \right)}  +\morder{\frac{Q_a^2 }{p^2} \log^2{\left( \frac{p^2}{\mu^2}\right)}}\right]  \, .
\label{IS-num}
\eea
The second equality above follows from the fact that the integral of each term in the first equality is of the form of \eq{In-A}, and thus we used the expressions \eq{I2-A}--\eq{I3-A}. Inserting \eq{IS-num} and \eq{Stilde-final-A} into \eq{spec-ind-IS-reg} we arrive at
\be
\left. z \frac{dI}{dz} \right|_{\rm IS} = 2\frac{\alpha_s C_R}{\pi} \left. \left\{ \log{\left( \frac{p^2}{C e^2 \mu^2} \right)} +\morder{\frac{Q_a^2 }{p^2} \log^2{\left( \frac{p^2}{\mu^2}\right)}} \right\} \right|_{\rm ind} \, .
\ee
In the latter expression, the first term in the bracket is independent of $Q_a$ and thus cancels when taking the difference between a nucleus and a proton target, as specified by \eq{ind-prescription}. Thus, the contribution from purely initial (final) state radiation is of order 
\be
\label{IS-magnitude}
\left. z \frac{dI}{dz} \right|_{\rm IS} \sim  \alpha_s \, \morder{\frac{{Q}_{a}^2 }{p^2} \log^2{\left( \frac{p^2}{\mu^2}\right)} }\, ,
\ee
where we assumed ${Q}_{a} \gg {Q}_{a {\rm p}}$ (with $Q_{a {\rm p}}$ the saturation scale of the parton $a$ in the proton, see footnote~\ref{foot:sat-def}). 

\subsection{Contribution from the interference term}
\label{app:int}

Here we present a more rigorous derivation of the interference contribution \eq{spec-ind-INT-1} than that presented in section~\ref{sec:int}.  

First, we perform the change of variable $\rvec \to \rvec - z{\xvec}$ (suggested by the discussion of section~\ref{sec:int}) in the numerator of \eq{spec-ind-INT-1} to obtain (within the limit $z \ll  1$)
\be
\label{spec-ind-INT-reg}
\hskip -3mm  \left. z \frac{dI}{dz} \right|_{\rm INT} = \frac{2 C_R \alpha_s}{\pi^2} \left. \frac{\int_{_\rvec}  e^{i \pvec \cdot \rvec} \int_{\xvec} \, \frac{{\xvec} \cdot ({\xvec} + \rvec)}{{\xvec}^2 ({\xvec} + \rvec)^2} \, e^{-i z \pvec \cdot \xvec} \left\{ 1 -  \left[S_a({\xvec}+\rvec) \, S_a({\xvec})\right]^{\sigma} S_a(\rvec)^{\rho} \right\} }{\int_{_\rvec} \, e^{i \pvec \cdot \rvec} S_a(\rvec)} \right|_{\rm ind} \! \! . 
\ee
The latter can be split into two mathematically well-defined terms:
\bea
\label{spec-ind-INT-split}
&& \hskip 3cm \left. z \frac{dI}{dz} \right|_{\rm INT} = \left. z \frac{dI}{dz} \right|_{\rm INT, 1} + \left. z \frac{dI}{dz} \right|_{\rm INT, 2}  \\ \nn \\
&& \hskip -10mm  \left. z \frac{dI}{dz} \right|_{\rm INT, 1} = \frac{2 C_R \alpha_s}{\pi^2} \left. \frac{\int_{_\rvec}  e^{i \pvec \cdot \rvec} \int_{\xvec} \, \frac{{\xvec} \cdot ({\xvec} + \rvec)}{{\xvec}^2 ({\xvec} + \rvec)^2} \, e^{-i z \pvec \cdot \xvec} \left\{ 1 -  [S_a({\xvec}+\rvec)\, S_a({\xvec})]^{\sigma} \right\} }{\int_{_\rvec} \, e^{i \pvec \cdot \rvec} S_a(\rvec)} \right|_{\rm ind}  \label{cont1} \\ 
&&  \hskip -10mm \left. z \frac{dI}{dz} \right|_{\rm INT, 2} = \frac{2 C_R \alpha_s}{\pi^2} \left. \frac{\int_{_\rvec}  e^{i \pvec \cdot \rvec} \left[ 1 -  S_a(\rvec)^{\rho} \right] \int_{\xvec} \, \frac{{\xvec} \cdot ({\xvec} + \rvec)}{{\xvec}^2 ({\xvec} + \rvec)^2} \, e^{-i z \pvec \cdot \xvec} \, [S_a({\xvec}+\rvec) \, S_a({\xvec})]^{\sigma} }{\int_{_\rvec} \, e^{i \pvec \cdot \rvec} S_a(\rvec)} \right|_{\rm ind}  \label{cont2} 
\eea

\subsubsection*{contribution \eq{cont1}}

After the change of variable $\rvec \to \rvec - \xvec$, the numerator of \eq{cont1} becomes (when $z \ll 1$, see footnote~\ref{foot:smallz})
\be
\label{cont1-num}
\int_{_\rvec}  e^{i \pvec \cdot \rvec}  \frac{\rvec}{\rvec^2} \int_{_\xvec}  e^{-i \pvec \cdot \xvec}  \frac{\xvec}{\xvec^2}  - \int_{_\rvec}  e^{i \pvec \cdot \rvec}  \frac{\rvec}{\rvec^2} \, S_a(\rvec)^{\sigma} \int_{_\xvec}  e^{-i \pvec \cdot \xvec}  \frac{\xvec}{\xvec^2} \, S_a({\xvec})^{\sigma} \, .
\ee 
When $\pvec$ is large, one can expand both $S_a(\rvec)$ and $S_a(\xvec)$ given by \eq{S-small-x} at small $\rvec$ and small $\xvec$. Using the identity \eq{conformal-identity}, a straightforward calculation shows that \eq{cont1-num} is of the form
\be
{\rm \eq{cont1-num}} \propto \frac{Q_a^2}{p^4} \left[ 1+ \morder{\frac{Q_a^2 }{p^2} \log{\left( \frac{p^2}{\mu^2}\right)} } \right] \, .
\ee
Dividing the latter by \eq{Stilde-final-A}, we find that the contribution \eq{cont1} is of order (assuming ${Q}_{a} \gg {Q}_{a {\rm p}}$)
\be
\label{cont1-final}
\left. z \frac{dI}{dz} \right|_{\rm INT, 1} \sim \alpha_s \left[ 1+ \morder{\frac{Q_a^2 }{p^2} \log{\left( \frac{p^2}{\mu^2}\right)} } \right]_{\rm ind} \sim \alpha_s \, \morder{\frac{{Q}_{a}^2 }{p^2} \log{\left( \frac{p^2}{\mu^2}\right)}} \, .
\ee
Up to logarithms, \eq{cont1-final} is of the same order as the contribution \eq{IS-magnitude} from purely initial state radiation. 

\subsubsection*{contribution \eq{cont2}}

The study of \eq{cont2}, which will bring the dominant contribution to the induced coherent spectrum, is somewhat simplified by performing the additional change of variable $\xvec \to \xvec - {\rvec}/{2}$ in \eq{cont2}, yielding\footnote{Under this change of variable the phase $\pvec \cdot \rvec$ of the factor $e^{i \pvec \cdot \rvec}$ acquires a phase shift $\frac{z}{2} \, \pvec \cdot \rvec$, which can however be neglected compared to $\pvec \cdot \rvec$ in the $z \ll 1$ limit, see footnote~\ref{foot:smallz}.}
\be
\label{cont2-shifted}
\left. z \frac{dI}{dz} \right|_{\rm INT, 2} = \frac{2 C_R \alpha_s}{\pi^2} \left. \frac{\int_{_\rvec}  e^{i \pvec \cdot \rvec} \left[ 1 -  S_a(\rvec)^{\rho} \right] \int_{\xvec} \, \frac{{\xvec}^2 - {\rvec^2}/{4}}{(\xvec - \frac{\rvec}{2})^2 (\xvec + \frac{\rvec}{2})^2 } \, e^{-i z \pvec \cdot \xvec} \, [S_a({\xvec}+\frac{\rvec}{2}) \, S_a({\xvec}-\frac{\rvec}{2})]^{\sigma} }{\int_{_\rvec} \, e^{i \pvec \cdot \rvec} S_a(\rvec)} \right|_{\rm ind} 
\ee

When $\pvec$ is large, the denominator of \eq{cont2-shifted} is expressed using \eq{Stilde-final-A}, and in the numerator of \eq{cont2-shifted} one can use
\be
\label{Sa-rho}
1 -  S_a(\rvec)^{\rho} \simeq  \rho\, \frac{Q_a^2 \, r^2}{8} \log{\left( \frac{1}{r^2 \mu^2} \right)} \left[ 1 + \morder{Q_a^2\, r^2 \log(\mu r)} \right] \, .
\ee

Let us neglect for the moment the correction of relative order $\sim \morder{Q_a^2 \, r^2\log(\mu r)}$ in \eq{Sa-rho}. An overall factor $Q_a^2$ cancels between numerator and denominator in \eq{cont2-shifted}, and only the factor $[S_a({\xvec}+\frac{\rvec}{2}) \, S_a({\xvec}-\frac{\rvec}{2})]^{\sigma}$ in the numerator of \eq{cont2-shifted} depends on $Q_a^2$. We can thus rewrite \eq{cont2-shifted} by applying the {\it induced prescription} \eq{ind-prescription} to this part only,
\be
\label{cont2-bis}
\hskip -3mm
\left. z \frac{dI}{dz} \right|_{\rm INT, 2} = \frac{2 C_R \alpha_s}{\pi^2} \frac{\int_{_\rvec}  e^{i \pvec \cdot \rvec} \rho\,  r^2  \log{(\mu r)} \int_{\xvec} \, \frac{{\xvec}^2 - {\rvec^2}/{4}}{(\xvec - \frac{\rvec}{2})^2 (\xvec + \frac{\rvec}{2})^2 }  \, e^{-i z \pvec \cdot \xvec} \, [S_a({\xvec}+\frac{\rvec}{2})^\sigma \, S_a({\xvec}-\frac{\rvec}{2})^{\sigma} ]_{\rm ind}}{\int_{_\rvec} \, e^{i \pvec \cdot \rvec}  (-r^2)  \log{(\mu r)}} \, ,
\ee
where $\left[ S_a({\xvec}+\frac{\rvec}{2})^\sigma \, S_a({\xvec}-\frac{\rvec}{2})^{\sigma} \right]_{\rm ind} \equiv  S_{a}({\xvec}+\frac{\rvec}{2})^\sigma \, S_{a}({\xvec}-\frac{\rvec}{2})^{\sigma} -  S_{a{\rm p}}({\xvec}+\frac{\rvec}{2})^\sigma \, S_{a{\rm p}}({\xvec}-\frac{\rvec}{2})^{\sigma}$.

The behavior of \eq{cont2-bis} at large $p$ can be obtained by expanding the integrand of the $\xvec$-integral at small $\rvec$. By setting $\rvec = \zerovec$ in the $\xvec$-integral, the leading term is safely obtained (the resulting integral is well-defined), namely
\be
\label{spec-ind-final-app}
\left. z \frac{dI}{dz} \right|_{\rm INT, 2} =  (2 C_R -C_A) \, \frac{\alpha_s}{\pi^2} \int \frac{d^2\xvec}{\xvec^2} \, e^{-i z \pvec \cdot \xvec} \, \left[- S_g({\xvec}) \right]_{\rm ind} 
\, ,
\ee
where we used $2C_R \, \rho = 2 C_R -C_A$, $S_a({\xvec})^{2\sigma} = S_a({\xvec})^{\frac{C_A}{C_R}}=S_g({\xvec})$, and $\left[- S_g({\xvec}) \right]_{\rm ind} = S_{g{\rm p}}({\xvec}) - S_{g}({\xvec})$. The expression \eq{spec-ind-final-app} coincides with \eq{spec-master} found in section~\ref{sec:int}. The subleading term of \eq{cont2-bis} in the large $p$ limit is obtained by keeping the $\morder{r^2}$ term in the integrand of the $\xvec$-integral. A straightforward (though somewhat cumbersome) calculation shows that the contribution of this term to the induced spectrum can be bounded by $\morder{\alpha_s \, ({Q}_{a}^2/{p^2}) \log^2{\left( {p^2}/{\mu^2}\right)}}$, and is thus of the same order (up to logarithms) as the terms neglected until now. 

Finally, we estimate the magnitude of the correction to \eq{spec-ind-final-app} brought by the term $\sim \morder{Q_a^2 \, r^2 \log(\mu r)}$ in \eq{Sa-rho}. This term gives a contribution to the induced spectrum of the form
\be
\label{correction-order}
\left. z \frac{dI}{dz} \right|_{\rm corr.} \sim \alpha_s \left. \frac{\int_{_\rvec}  e^{i \pvec \cdot \rvec} \, Q_a^2 \, r^4  \log^2(\mu r) \int_{\xvec} \, \frac{{\xvec}^2 - {\rvec^2}/{4}}{(\xvec - \frac{\rvec}{2})^2 (\xvec + \frac{\rvec}{2})^2 } \, e^{-i z \pvec \cdot \xvec} \, [S_a({\xvec}+\frac{\rvec}{2}) \, S_a({\xvec}-\frac{\rvec}{2})]^{\sigma} }{\int_{_\rvec} \, e^{i \pvec \cdot \rvec} \,  r^2  \log(\mu r)}  \right|_{\rm ind} \, .
\ee
It is easy to see that the $\xvec$-integral in the numerator of \eq{correction-order} is dominated by the logarithmic domain $r \sim 1/p \ll  x \ll {\rm min}(\frac{1}{Q_a},\frac{1}{zp})$. Thus, up to logarithmic factors, \eq{correction-order} is of order
\be
\label{correction-order-2}
\left. z \frac{dI}{dz} \right|_{\rm corr.} \sim \alpha_s \left. \frac{(Q_a^2 / p^6)}{(1/p^4)}  \right|_{\rm ind}  \sim \alpha_s \, \morder{\frac{{Q}_{a}^2 }{p^2}} \, .
\ee

In summary, the interference contribution is given by \eq{spec-ind-final-app}, up to terms at most on the order of \eq{IS-magnitude}.


\providecommand{\href}[2]{#2}\begingroup\raggedright
\endgroup

\end{document}